 \documentclass[final,5p,times,twocolumn]{elsarticle}

\usepackage{latexsym}
\usepackage{graphicx}  

\usepackage{subfigure}
\usepackage[T1]{fontenc}
\usepackage[latin1]{inputenc}
\usepackage{dsfont}
\usepackage{amssymb}
\usepackage{amsmath}
\usepackage{amsfonts}
\usepackage{latexsym}
\usepackage{color}

\usepackage{nomencl}
\makenomenclature

\journal{Automatica}

\begin{document}

\begin{frontmatter}

%\title{{Are hybrid control systems (satisfying hybrid basic conditions) really robust to measurement noise?}}
\title{On the robustness of hybrid control systems\\ to measurement noise and actuator disturbances\tnoteref{dpi}}

\author[a1]{Alfonso~Ba\~nos\corref{cor1}} 
\ead{abanos@um.es}

\author[a2]{Miguel A. Dav\'o}
\ead{Miguel.Davo-Navarro@gipsa-lab.fr}

\author[a1]{Cristian D. C\'anovas}
\ead{cristiandavid.canovas@um.es}

 \tnotetext[dpi]{This work has been supported by {\em FEDER-EU} and {\em Ministerio de Ciencia e Innovaci\'on (Gobierno de Espa\~na)} under project DPI2016-79278-C2-1-R. %, and by the ANR project LimICoS contract number 12-BS03-005-01}% <-this % stops a space
}

\cortext[cor1]{Corresponding author}

\address[a1]{Universidad de Murcia, Dept. Inform\'atica y Sistemas, 30100 Murcia, Spain}

\address[a2]{Université Grenoble Alpes, CNRS, GIPSA-Lab, F-38000 Grenoble, France}

%\address[a3]{Dept. Inform\'atica y Sistemas, Univ. of Murcia, 30100 Murcia, Spain}

%%%%%%%%%%%%%%%%%%%%%%%%%%%%%%%%%%%%%%%%%%%%%%%%%%%%%%%%%%%%%%%%%%%%%%%%%%%%%%%%
\begin{abstract}
\textcolor{black}{Robustness of hybrid control systems to measurement noise, actuator disturbances, and more generally perturbations, is analyzed. The relationship between the robustness of a hybrid control system and of its implementations is emphasized. \textcolor{black}{Firstly, a formal definition of  implementation of a hybrid control system is provided, based on the uniqueness of the solutions.} Then, two examples are analyzed in detail, showing how the previously developed robustness property fails to guarantee that the implementations, necessarily used in control practice, are also robust. A new concept of strong robustness is proposed, \textcolor{black}{which guarantees that at least jumping-first and flowing-first implementations are robust when the hybrid control system is strongly robust. In addition, we provide a sufficient condition for strong robustness based on the previously developed hybrid relaxation results.}}
\end{abstract}
%%%%%%%%%%%%%%%%%%%%%%%%%%%%%%%%%%%%%%%%%%%%%%%%%%%%%%%%%%%%%%%%%%%%%%%%%%%%%%%%


\end{frontmatter}

{\em Notation}: 
$\mathds{R}_{\geq 0}$ is the set of non-negative real numbers, $\mathds{R}^n$ is the $n$-dimensional Euclidean space, and $\mathbf{x} = (x_1,\cdots,x_n) \in \mathds{R}^n$ is a column vector; $\|\mathbf{x}\|$ is the euclidean norm.  $\mathds{B}$ is the closed unit ball in $\mathds{R}^n$ centered at the origin. For a set $K \subset \mathds{R}^n$, $\text{con}(K)$ denotes the convex hull of $K$, $\overline{K}$ is its closure, and $\text{int }K$ is the interior of $K$.  $\mathcal{S}_{\mathcal H}(\xi)$ is the set of maximal solutions $\phi$ to the hybrid system ${\mathcal H}$ with $\phi(0,0) =\xi$. $\text{dom}$ stands for domain, and $\setminus $ denotes sets difference.

\section{INTRODUCTION}
This work is focused on the hybrid system framework developed in \cite{GSTbook} (and references therein), that following \cite{Schutter09}, will be referred to as Hybrid Inclusions (HI) framework.  The reader is referred to \cite{GSTbook,Goebel09} for a detailed exposition of the HI framework. For the sake of completeness, a minimal background is % the definition of solution and the hybrid basic conditions are 
given in the appendices.
In particular, the work is centered in hybrid control systems that may be modeled by hybrid systems with the following data in $\mathds{R}^n$: i) the {\em flow set} $\mathcal{C}\subset \mathds{R}^n$, ii) the {\em flow mapping} $f: \mathds{R}^n \rightarrow \mathds{R}^{n}$, 
iii) the {\em jump set} ${\mathcal D} \subset \mathds{R}^n$, and iv) the {\em jump mapping} $g:\mathds{R}^n \rightarrow \mathds{R}^{n}$. A hybrid system with these data is represented as ${\mathcal H} = ({\mathcal C},f,{\mathcal D},g)$ and is given by 
\begin{equation}
\mathcal{H}: \left\{  \begin{array}{llll}
\mathbf{\dot{x}}=f(\mathbf{x}), \quad & \mathbf{x}  \in \mathcal{C}, \\
\mathbf{x}^+=g(\mathbf{x}), \quad & \mathbf{x} \in \mathcal{D}.  
\end{array}
\right.
\label{eq-H}
\end{equation}

For ${\mathcal H}$, the so-called {\em hybrid basic conditions} defined in \cite{GSTbook}  (see also App. B) are trivially satisfied if $\mathcal{C}$ and $\mathcal{D}$ are both closed subsets of $\mathds{R}^{n}$, and for example, $f$ and $g$ are continuous functions. On the other hand, a perturbation of ${\mathcal H}$ is a family of hybrid systems ${\mathcal H}_\delta$ with data $(\mathcal{C}_\delta,f_\delta,\mathcal{D}_\delta,g_\delta)$, and perturbation parameter $\delta>0$. %For ${\mathcal H}_\delta$, the {\em convergence property} is defined in \cite{goebel06}. 

%The reader is referred to \cite{GSTbook,Goebel09} for a detailed exposition to the HI framework. For the sake of completeness, a minimal background is % the definition of solution and the hybrid basic conditions are given in the appendices.

%Robustness of $\mathcal{H}$ to perturbations is developed in \cite{goebel06}, being this related with the closeness of solutions to $\mathcal{H}$ and solutions to $\mathcal{H}_\delta$ for vanishing values of the parameter $\delta$.  Although the robustness property is not explicitly defined %, in fact the property is not explicitly named
%in \cite{goebel06}, and named as "dependence on initial conditions and perturbations" in (\cite{GSTbook}, Prop. 6.34), %it is presumably based on the following  
%it can be made explicit by the following definition (extracted from \cite{goebel06}- Corollary 5.5): {\em for a compact set $K \subset \mathds{R}^n$ such that $\mathcal{H}$ is forward complete from $K$, the hybrid system $\mathcal{H}$ is robust to a perturbation given by $\mathcal{H}_\delta$  if for any $\epsilon >0$ and $(T,J) \in \mathds{R}_{\geq 0}\times \mathds{N}$ there exists $\delta^\star >0$ with the following property: for any $\delta \in (0,\delta^\star]$ and any $\mathbf{x}_\delta \in \mathcal{S}_{\mathcal{H}_\delta}(K+\delta \mathds{B})$ there exists a solution $\mathbf{x}$ to ${\mathcal H}$, with $\mathbf{x}(0,0) \in K$ such that $\mathbf{x}_\delta$ and $\mathbf{x}$ are $(T,J,\epsilon)$-close.
%} Note that this property includes a particular notion of continuous dependence on the initial condition.

\textcolor{black}{
Robustness of $\mathcal{H}$ to perturbations is developed in \cite{goebel06}, being this related with the closeness of solutions to $\mathcal{H}$ and solutions to $\mathcal{H}_\delta$ for small enough values of the parameter $\delta$.  %Although the robustness property is not explicitly defined, in fact the property is not explicitly named, and named as "dependence on initial conditions and perturbations" in (\cite{GSTbook}, Prop. 6.34), an implicit definition can be obtained from \cite{goebel06}- Corollary 5.5. 
Although an explicit definition of the robustness property has not been given, the property is implicitly defined in \cite{goebel06} (see Corollary 5.5), and in (\cite{GSTbook}, Prop. 6.34) named as "dependence on initial conditions and perturbation".  
%it is presumably based on the following  
%it can be made explicit by the following definition (extracted from \cite{goebel06}- Corollary 5.5): {\em for a compact set $K \subset \mathds{R}^n$ such that $\mathcal{H}$ is forward complete from $K$, the hybrid system $\mathcal{H}$ is robust to a perturbation given by $\mathcal{H}_\delta$  if for any $\epsilon >0$ and $(T,J) \in \mathds{R}_{\geq 0}\times \mathds{N}$ there exists $\delta^\star >0$ with the following property: for any $\delta \in (0,\delta^\star]$ and any $\mathbf{x}_\delta \in \mathcal{S}_{\mathcal{H}_\delta}(K+\delta \mathds{B})$ there exists a solution $\mathbf{x}$ to ${\mathcal H}$, with $\mathbf{x}(0,0) \in K$ such that $\mathbf{x}_\delta$ and $\mathbf{x}$ are $(T,J,\epsilon)$-close.
%} Note that this property includes a particular notion of continuous dependence on the initial condition.
%In the rest of this work, we analyze robustness of hybrid control systems to measurement noise. 
More specifically, for hybrid control systems, measurement noise is embedded in a type of  perturbation referred to as outer perturbation  (\cite{Goebel09}). It is known that the recent work \cite{copp16} gives a related definition of robustness to measurement noise in an implicit way (Theorem 5.5), for a specific type of hybrid systems.
%, and robustness to measurement noise is approached by robustness to that perturbation. 
}

In the HI framework, robustness to perturbations is usually approached by imposing the hybrid basic conditions on $\mathcal{H}$, and the {\em convergence property} on its perturbation $\mathcal{H}_\delta$ (\cite{goebel06,GSTbook,Goebel09}). This results in a regularization of $\mathcal{H}$ that usually leads to a non-deterministic system, in the sense that there may exist several solutions to $\mathcal{H}$ from some initial points. 

However, when $\mathcal{H}$ is a feedback control system, and specially when the hybrid dynamics is determined by the feedback controller, its practical implementation entails a decision mechanism such that a unique solution is selected within all the possible solutions.  A key question in control practice is whether the different implementations of a robust hybrid control system keep the robustness property or not. This is a non-trivial question that apparently has not been previously approached.  %in the HI framework. 
The main goal of this work is to analyze this question, 
%the robustness of hybrid control systems to measurement noise and external disturbances
and specifically to investigate the relation between the robustness of a hybrid control system and of its implementations.

% Using the HI framework, the property is posed as a particular case of robustness to perturbations.
\textcolor{black}{
In Section 2, the hybrid control system setup, including measurement noise and actuator disturbances as exogenous signals, is given. The model is embedded in a more general hybrid system with perturbations. Then, robustness to perturbations is formally defined by using a definition similar to the implicit concept given in (\cite{goebel06}, Corollary 5.5), and a basic HI result is recalled.
% a definition similar to the above definition of robustness to perturbations, and a basic HI result is recalled. %, which is that a hybrid system is robust to measurement error if it satisfies the basic hybrid conditions. 
In Section 3, we define the concept of implementation of a hybrid control system;  %based on two basic properties: the solutions of an implementation should be unique and the hybrid system and its implementation should share the state-space and the set of initial condition. 
in addition, we use two examples %to analyze the robustness property,
to show that robustness of a hybrid control system does not guarantee the existence of robust implementations. Finally, Section 4 introduces the new  property of strong robustness to perturbations, that leads to robustness of implementations; moreover, some hybrid relaxation results are used to derive sufficient conditions for strong robustness. Some concluding remarks are also elaborated. 
}

\section{Preliminaries and background}
%In the HI framework, robustness to measurement noise is developed as a particular case of robustness to  perturbations. Although this property has been more or less explicitly analyzed in several works (for example in \cite{Goebel09} noise is explicitly embedded in a more general outer perturbation), to the authors knowledge a formal definition of the property is missing. In the following, we define the robustness to measurement noise in the spirit of the HI framework with the goal of developing a precise analysis in the next section. 
\begin{figure}[t] %  figure placement: here, top, bottom, or page
  % \small
   \centering
    \includegraphics[width=4.15cm]{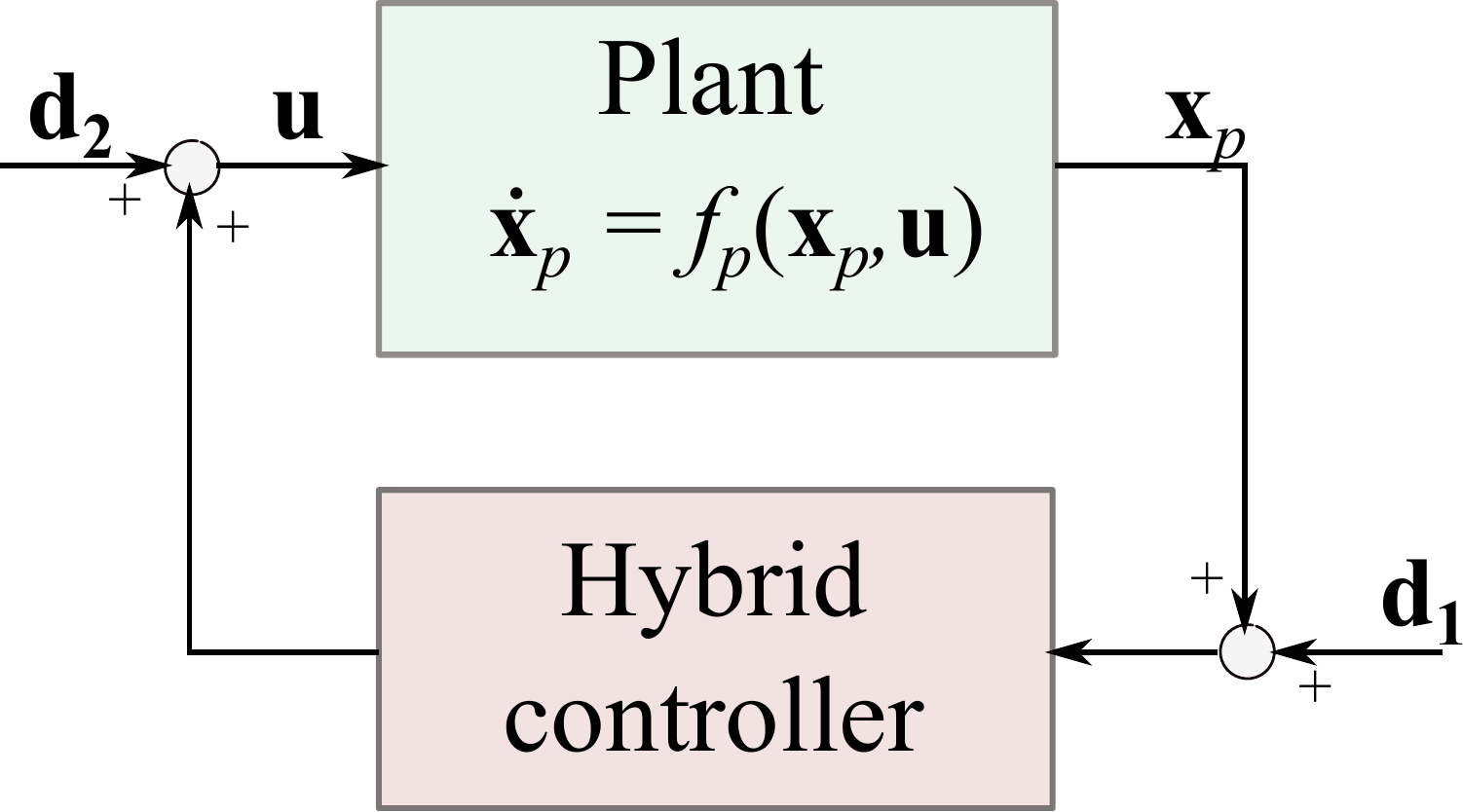}
    \caption{A hybrid control system, with a continuous-time plant (with state $\mathbf{x}_p$) and a feedback hybrid controller (with state $\mathbf{x}_c$). The feedback loop is perturbed by errors both in the measurement (noise $\mathbf{d}_1$) and the actuator (disturbance $\mathbf{d}_2$). %The perturbed closed-loop hybrid system $\mathcal{H}_{(\mathbf{d}_1,\mathbf{d}_2)}^\text{cl}$ is given by (5), while the (unperturbed) hybrid control system $\mathcal{H}^\text{cl}$ is obtained by making $\mathbf{d}_1 = \mathbf{d_2} = \mathbf{0}$ in (5). Note these hybrid systems models are particular cases of $\mathcal{H}_{(\mathbf{e}_1,\mathbf{e}_2,\mathbf{e}_3)}$ and $\mathcal{H}$, as given by \eqref{eq-He} and \eqref{eq-H}, respectively.
    }
    \label{fig-controlsetup}
\end{figure}

In this work, the main focus is on hybrid control systems that can be modeled by \eqref{eq-H}. This is, for example, the case in which a continuous-time plant is controlled by a hybrid controller (see Fig. \ref{fig-controlsetup}), a type of control  which appears in a broad class of industrial applications (\cite{Goebel09}). The plant is described by the differential equation: 
\begin{equation}
\dot{\mathbf{x}}_p = f_p(\mathbf{x}_p,\mathbf{u})
\end{equation}
where $\mathbf{x}\in \mathds{R}^{n_p}$, $\mathbf{u} \in \mathds{R}^{n_r}$, and $f_p$ is continuous. The hybrid controller, with state $\mathbf{x}_c\in \mathds{R}^{n_c}$, is defined by a flow set $\mathcal{C}\subset \mathds{R}^{n_p+n_c}$, a flow map $f_c:\mathds{R}^{n_p+n_c} \rightarrow \mathds{R}^{n_c}$, a jump set $\mathcal{D}\subset \mathds{R}^{n_p+n_c}$, a jump map $g_c:\mathds{R}^{n_p+n_c} \rightarrow \mathds{R}^{n_c}$, and a feedback control law $k_c:\mathds{R}^{n_p+n_c} \rightarrow \mathds{R}^{n_r}$ that specifies the control signal $\mathbf{u}$. Defining the state of the closed-loop system as $\mathbf{x}= (\mathbf{x}_p,\mathbf{x}_c)$ and $n = n_p+n_c$, it results that the closed-loop system is a hybrid system $\mathcal{H}^\text{cl} = (\mathcal{C},f,\mathcal{D},g)$ as given by \eqref{eq-H} with flow map $f:\mathds{R}^{n} \rightarrow \mathds{R}^{n}$ defined as
\begin{equation}
f(\mathbf{x}_p, \mathbf{x}_c) = 
\left(
\begin{array}{ccc}
  f_p(\mathbf{x}_p, k_c(\mathbf{x}_p,\mathbf{x}_c))    \\
     f_c(\mathbf{x}_p,\mathbf{x}_c)
\end{array}
\right)
\end{equation}
and jump map $g:\mathds{R}^{n} \rightarrow \mathds{R}^{n}$ defined as
\begin{equation}
g(\mathbf{x}_p, \mathbf{x}_c) = 
\left(
\begin{array}{ccc}
  \mathbf{x}_p    \\
     g_c(\mathbf{x}_p,\mathbf{x}_c)
\end{array}
\right)
\end{equation}
The main goal is to analyze the robustness properties of this hybrid control system with respect to measurement noise, and more generally with respect to external disturbances. Considering the measurement noise $\mathbf{d}_1 \in \mathds{R}^{n_p}$ and the actuator disturbance $\mathbf{d}_2\in \mathds{R}^{n_r}$ as perturbations (Fig. \ref{fig-controlsetup}), the perturbed hybrid control system $\mathcal{H}_{(\mathbf{d}_1,\mathbf{d}_2)}^\text{cl}$ is given by: 
% as external perturbations (Fig. \ref{fig-controlsetup}) the measurement noise $\mathbf{d}_1 \in \mathds{R}^{n_p}$ and the actuator perturbation $\mathbf{d}_2\in \mathds{R}^{n_r}$.  The perturbed hybrid control system $\mathcal{H}_{(\mathbf{d}_1,\mathbf{d}_2)}^\text{cl}$ is given by: 
\begin{equation}
\mathcal{H}^\text{cl}_{(\mathbf{d}_1,\mathbf{d}_2)}:\left\{  \begin{array}{llll}
\mathbf{\dot{x}}=\left(
\begin{array}{ccc}
  f_p(\mathbf{x}_p, k_c(\mathbf{x}_p+\mathbf{d}_1,\mathbf{x}_c)+\mathbf{d}_2)    \\
     f_c(\mathbf{x}_p + \mathbf{d}_1,\mathbf{x}_c)
\end{array}
\right)
, &(\mathbf{x}_p + \mathbf{d}_1, \mathbf{x}_c) \in \mathcal{C}, \\
\mathbf{x}^+= 
\left(
\begin{array}{ccc}
  \mathbf{x}_p    \\
     g_c(\mathbf{x}_p + \mathbf{d}_1,\mathbf{x}_c)
\end{array}
\right),&(\mathbf{x}_p+ \mathbf{d}_1, \mathbf{x}_c) \in \mathcal{D}.  
\end{array}
\right.
\label{eq-Hp}
\end{equation}
In order to embed this perturbed hybrid control system in a more general perturbation, let us consider perturbations  $\mathbf{e}_1$, $\mathbf{e}_2$, $\mathbf{e}_3 \in \mathds{R}^n$ and the following perturbed hybrid system
%Thus, the perturbed hybrid control system \eqref{eq-Hp} is simply given b
\begin{equation}
{\mathcal{H}_{(\mathbf{e}_1,\mathbf{e}_2,\mathbf{e}_3)}}:\left\{  \begin{array}{lcrlcr}
\mathbf{\dot{x}} = f (\mathbf{x}+ \mathbf{e}_1)+ \mathbf{e}_2, \quad & \mathbf{x}+ \mathbf{e}_1\in \mathcal{C}, \\
\mathbf{x}^+ = g(\mathbf{x} + \mathbf{e}_1) + \mathbf{e}_3, \quad & \mathbf{x}+ \mathbf{e}_1 \in \mathcal{D}. %\\
%\mathbf{x}(0) &= \xi 
\end{array}
\right. 
\label{eq-He}
\end{equation}
Then, the perturbed hybrid control system $\mathcal{H}_{(\mathbf{d}_1,\mathbf{d}_2)}^\text{cl}$ is embedded in the general system \eqref{eq-He} by simply considering the following perturbations 
%Now, consider the {\em error} signals $\mathbf{e}_1, \mathbf{e}_2, \mathbf{e}_3 \in \mathbf{R}^{n}$ defined, in this case, as 
 \begin{equation}                        
\begin{array}{ll}
  \mathbf{e}_1 = (\mathbf{d}_1,\mathbf{0}) \\
\mathbf{e}_2 = (f_p(\mathbf{x}_p, k_c(\mathbf{x}_p+\mathbf{d}_1,\mathbf{x}_c)+\mathbf{d}_2) - f_p(\mathbf{x}_p + \mathbf{d}_1, k_c(\mathbf{x}_p+\mathbf{d}_1,\mathbf{x}_c)), \mathbf{0}) \\
\mathbf{e}_3 = - (\mathbf{d}_1,\mathbf{0}), 
\end{array}
\label{eq-e}
\end{equation}
where it directly follows that $\|\mathbf{e}_1\|$ and $\|\mathbf{e}_3\|$ are arbitrarily small, and also $\|\mathbf{e}_2\|$ by continuity of $f_p$,  when $\|\mathbf{d}_1\|$ and  $\|\mathbf{d}_2\|$ are small enough\footnote{Note that since $f_p$ is continuous, then for any $\varepsilon > 0$ there exists $\delta>0$ such as if $\|(\mathbf{x}_p, k_c(\mathbf{x}_p+\mathbf{d}_1,\mathbf{x}_c)+\mathbf{d}_2) - (\mathbf{x}_p + \mathbf{d}_1, k_c(\mathbf{x}_p+\mathbf{d}_1,\mathbf{x}_c)\| = \| (-\mathbf{d}_1,\mathbf{d}_2)\| < \delta$ then $\|\mathbf{e}_2\| < \epsilon$. This is obtained for example for $\|\mathbf{d}_1\| < \min \{\delta/2,\varepsilon/2\}$ and  $\|\mathbf{d}_2\| < \min \{\delta/2,\varepsilon/2\}$, which on the other hand directly makes $\|\mathbf{e}_1\| < \varepsilon$ and $\|\mathbf{e}_3\| < \varepsilon$. }. 
%Thus, the perturbed hybrid control system \eqref{eq-Hp} is simply given by 
%\begin{equation}
%{\mathcal{H}_{(\mathbf{e}_1,\mathbf{e}_2,\mathbf{e}_3)}}:\left\{  \begin{array}{lcrlcr}
%\mathbf{\dot{x}} = f (\mathbf{x}+ \mathbf{e}_1)+ \mathbf{e}_2, \quad & \mathbf{x}+ \mathbf{e}_1\in \mathcal{C}, \\
%\mathbf{x}^+ = g(\mathbf{x} + \mathbf{e}_1) + \mathbf{e}_3, \quad & \mathbf{x}+ \mathbf{e}_1 \in \mathcal{D}. %\\
%%\mathbf{x}(0) &= \xi 
%%\end{array}
%\right. 
%\label{eq-He}
%\end{equation}

The perturbed hybrid system \eqref{eq-He} is considered in this work as a general perturbation model\footnote{A model similar to $\mathcal{H}_{(\mathbf{e}_1,\mathbf{e}_2,\mathbf{e}_3)}$ has been used to represent different hybrid feedback control systems with external perturbations due to measurement noise, actuator error, and other external disturbances (see \cite{Prieur01,PGT07,Goebel09}, and \cite{LS99} for the continuous-time systems case). Obviously, the model may also represent perturbed hybrid systems that are not necessarily feedback control systems.} 
of \eqref{eq-H} for arbitrary bounded perturbation signals $\mathbf{e}_i:\text{dom } \mathbf{e}_i \rightarrow \mathds{R}^n$, $i = 1,2,3.$ 
%\vspace{3cm}
%%%%%%
%%%%%%%%
%suppose that the measurements of the state $\mathbf{x}$ are noisy, and the noise is some bounded measurable signal $\mathbf{e}:\text{dom } \mathbf{e} \rightarrow \mathds{R}^n$.  If $\mathbf{x}+\mathbf{e}$ is the noisy state, then the hybrid system \eqref{eq-H} with measurement noise, denoted by $\mathcal{H}_\mathbf{e}$ and referred to as noisy hybrid system, is given by
%\begin{equation}
%\mathcal{H_{\mathbf{e}}}:\left\{  \begin{array}{lcrlcr}
%\mathbf{\dot{x}} = f (\mathbf{x}+ \mathbf{e}), \quad & \mathbf{x}+ \mathbf{e}\in \mathcal{C}, \\
%\mathbf{x}^+ = g(\mathbf{x} + \mathbf{e}), \quad & \mathbf{x}+ \mathbf{e} \in \mathcal{D}. %\\
%\mathbf{x}(0) &= \xi 
%\end{array}
%\right. 
%\label{eq-He}
%\end{equation}
Since the main goal is to analyze the impact of perturbations on solutions to  \eqref{eq-H} for small enough values of $\|\mathbf{e}_i\|$, $i = 1,2,3$, then a parameter $\delta$ is used and, for $i = 1,2,3$, $\mathbf{e}_i$ is simply defined as   $\mathbf{e}_i=\delta \mathbf{n}_i$, where $\delta >0$, $\mathbf{n}_ i$ is a measurable signal with $|\mathbf{n}_i(t,j)| \leq  1$ for any $(t,j) \in \text{dom } \mathbf{e}_i$; $\mathbf{n}_i$ will be referred to as an {\em admissible perturbation signal}.  In addition, for some admissible perturbation signals $\mathbf{n}_i$, $i =1,2,3$, %a family of perturbed hybrid systems 
$\mathcal{H}_{\delta}$ is defined as a perturbation of $\mathcal{H}$ given by 
\begin{equation}
\mathcal{H}_{\delta} = \{ \mathcal{H}_{(\mathbf{e}_1,\mathbf{e}_2,\mathbf{e}_3)}: \mathbf{e}_i=\delta \mathbf{n}_i, \delta >0, i = 1,2,3\} .
\label{eq-Hdeltan}
\end{equation}

\textcolor{black}{
Note that $\mathcal{H}_{\delta(\mathbf{n}_1,\mathbf{n}_2,\mathbf{n}_3)}$ corresponds to a hybrid system like \eqref{eq-He} with $(\mathbf{e}_1,\mathbf{e}_2,\mathbf{e}_3)= \delta(\mathbf{n}_1,\mathbf{n}_2,\mathbf{n}_3)$, and $\mathcal{H}_\delta$ is a set of hybrid systems. In some cases, perturbation signals only depends on time $t$ and not on $j$; then, the admissible perturbation signal $\mathbf{n}_i$ can be considered as given by setting $\mathbf{n}_i(t,j) = \mathbf{n}'_i(t)$ for all $(t,j) \in E$, for some function $\mathbf{n}'_i: \mathds{R}_{\geq 0} \rightarrow \mathds{R}$, and any arbitrary hybrid time domain $E$. 
}
%%%%%%%%%%%
%\vspace{0.25cm}

%In the HI framework, although the property of robustness to measurement noise (and external disturbances) has been more or less explicitly analyzed in several works (for example, in \cite{Goebel09} noise is explicitly embedded in a more general outer perturbation\footnote{A definition of outer perturbation is given in Example 5.3 of \cite{Goebel09}.}), to the authors knowledge a formal definition of the property is missing. 
In the following, we define robustness in the spirit of the HI framework with the goal of developing a precise analysis in the next section. 
\textcolor{black}{For the sake of simplicity, $\mathbf{n} = (\mathbf{n}_1,\mathbf{n}_2,\mathbf{n}_3)$, and thus  $\mathcal{H}_{\delta\mathbf{n}}=\mathcal{H}_{\delta(\mathbf{n}_1,\mathbf{n}_2,\mathbf{n}_3)}$, is used in the following. In addition, robustness to perturbations is used to denote robustness of the hybrid system \eqref{eq-H} to vanishing perturbations signals.}

\vspace{0.15cm}
{\bf Definition 2.1} (Robustness to perturbations) {\em For a compact set $K \subset \mathds{R}^n$ such that $\mathcal{H}$ is forward complete from $K$, 
%and an admissible noise signal $\mathbf{n}$,
 the hybrid system $\mathcal{H}$ is robust to perturbations if for any $\epsilon >0$ and $(T,J) \in \mathds{R}_{\geq 0}\times \mathds{N}$ there exists $\delta^\ast >0$ with the following property: for any admissible perturbation signal $\mathbf{n}$, any $\delta \in (0,\delta^\ast]$, and any $\mathbf{x}_{\delta \mathbf{n}} \in \mathcal{S}_{\mathcal{H}_{\delta \mathbf{n}}}(K+\delta \mathds{B})$ there exists a solution $\mathbf{x}$ to ${\mathcal H}$, with $\mathbf{x}(0,0) \in K$ such that $\mathbf{x}_{\delta \mathbf{n}}$ and $\mathbf{x}$ are $(T,J,\epsilon)$-close.
} 

\vspace{0.15cm}
Note that the definition of robustness includes a notion of uniformity with respect to continuous dependence on a set of initial conditions $K$. 
\textcolor{black}
{In particular, when the hybrid system is a hybrid control system $\mathcal{H}^\text{cl}$, robustness to perturbations means, in particular, robustness to the perturbation signals given by \eqref{eq-e}, then {robustness} to perturbations implies {\em robustness of the hybrid control system to measurement noise and actuator disturbances.}} %and a (normalized) noise signal $\mathbf{n}$. 

A significant result, that directly follows from Th. 5.4 and Corollary 5.5 in \cite{goebel06}, \textcolor{black}{embedding the perturbation \eqref{eq-Hdeltan} in a more general outer perturbation%(that satisfies the convergence property when hybrid basic conditions are met)
, }is stated in the following corollary. 

\vspace{0.15cm}
{\bf Corollary 2.2}  {\em A hybrid system $\mathcal{H}$ is robust to perturbations for some compact set $K\subset \mathds{R}^n$, where $\mathcal{H}$ is forward complete from $K$, if it satisfies the basic hybrid conditions.}
\section{Robustness of hybrid control system implementations}
Robustness to perturbations is a sound contribution of the HI framework, since it applies to hybrid systems with some simple properties (the hybrid basic conditions), and thus it may be applied to a generality of cases. In fact, the hybrid basic conditions are used as a mean to regularize hybrid systems and to equip them with the robustness to perturbation property (and some other useful properties like robust stability, see e.g. \cite{GSTbook}). \textcolor{black}{
The result is that, in general, hybrid systems are non-deterministic in the sense that several solutions may exist for a given initial point.
}

\textcolor{black}{
In control practice, the implementation of a hybrid control system $\mathcal{H}^\text{cl}$  (see Fig. 1)}%defined by \eqref{eq-Hcl}
, satisfying the hybrid basic conditions, entails a decision mechanism for the hybrid controller such that a unique solution is selected within all the theoretical possible solutions.  Such mechanism basically consists in choosing to jump or to flow at each instant in which both jumping and flowing are possible.

For the HI framework to be useful in control practice, we may expect that there exist implementations of the hybrid control system that inherits the robustness property; otherwise, any possible implementation will be sensitive to arbitrarily small measurement noise signals \textcolor{black}{
and/or actuator disturbances}.  Next, the notion of hybrid control system implementation is formalized; in addition, two examples are developed, showing that, in fact, the robustness property of hybrid control systems is not necessarily inherited by its implementations.%analyzing the robustness of hybrid systems and its implementations.}  

%In the rest of this work, the focus will be on hybrid control syst

%by simplicity $\mathcal{H}$ is us
\subsection{Implementation of a hybrid control system}

%Consider a hybrid control system $\mathcal{H}^\text{cl}$ %or more generally a hybrid system $\mathcal{H}$ like \eqref{eq-H} 
%that satisfies the hybrid basic conditions.
In the rest of this work, and with some abuse of notation, $\mathcal{H}$ will be indistinguishably used to denote a hybrid control system $\mathcal{H}^\text{cl}$ or more generally a hybrid system like \eqref{eq-H}. 
%\textcolor{black}{and has nontrivial solutions.} 
An {\em implementation} of $\mathcal{H}$ will be defined as a hybrid system $\mathcal{H}^I$ that has unique solutions, and those solutions are also solutions to $\mathcal{H}$. While $\mathcal{H}^I$ is an implementation of $\mathcal{H}$, we refer to the hybrid system $\mathcal{H}$ as an {\em abstraction} of $\mathcal{H}^I$.

\vspace{0.15cm}
{\bf Definition 3.1} (Implementation of a hybrid control system) {\em Consider a hybrid control system ${\mathcal H} = ({\mathcal C},f,{\mathcal D},g)$ that satisfies the hybrid basic conditions. A hybrid system ${\mathcal H}^I = ({\mathcal C}_I,f_I,{\mathcal D}_I,g_I)$ is an implementation of ${\mathcal H}$ if 
\begin{enumerate}
\item ${\mathcal C}_I \cup {\mathcal D}_I = {\mathcal C} \cup {\mathcal D}$;
\item for every $\xi \in \mathcal{C}_I \cup \mathcal{D}_I$, each solution $\phi \in \mathcal{S}_{\mathcal{H}^I}(\xi)$ is unique, and in addition, $\phi \in \mathcal{S}_{\mathcal{H}}(\xi)$.
\end{enumerate}
}

%Obviously, an {\em implementation of a hybrid control system} is defined is the same way substituting $\mathcal{H}$ by  $\mathcal{H}^\text{cl}$. 

Although this work is mainly focused on hybrid control system, the above definition is also valid for hybrid systems $\mathcal{H}$ like \eqref{eq-H} that are not necessarily hybrid control systems. 
The main motivations of the above definition are: firstly, to obtain implementations $\mathcal{H}^I$  which share the same state-space that its abstractions $\mathcal{H}$ (in this case $\mathds{R}^n$), and the same set of initial conditions $\mathcal{C} \cup \mathcal{D}$; and secondly, that implementations be deterministic hybrid systems in the sense that they have unique solutions for any initial point.

On the other hand, it directly follows from Def. 3.1 that $\text{int }\mathcal{C}_I \cap \text{int }\mathcal{D}_I = \varnothing$, otherwise ${\mathcal H}I$ would have several solutions $\phi$ with $\phi(0,0) \in \text{int }\mathcal{C}_I \cap \text{int }\mathcal{D}_I$. For those systems $\mathcal{H}$ such that $\text{int }\mathcal{C} \cap \text{int }\mathcal{D} = \varnothing$, condition 1 of Def. 3.1 would be directly guaranteed by considering implementations such that $\mathcal{C}=\bar{ \mathcal{C}_I}$ and  $\mathcal{D}=\bar{\mathcal{D}_I}$. In addition,  the abstraction $\mathcal{H}$ would be a Krasovskii regularization of $\mathcal{H}^I$ if $f_I=f$ and $g_I=g$.

%Finally, note that $f$ and $g$ are not set-valued functions, and thus, considering the hybrid basic conditions and the basic uniqueness conditions (see Appendix 1), it directly follows the following result on the existence of implementations.

%\vspace{0.15cm}
%{\bf Corollary III.2} There exists  an implementation $\mathcal{H}^I$ of a hybrid system $\mathcal{H}=(\mathcal{C},f,\mathcal{D},g)$,  satisfying the hybrid basic conditions, if the following conditions hold:
%\begin{enumerate}
%\item $\text{int }\mathcal{C} \cap \text{int }\mathcal{D} = \varnothing$;
%\item for every $\xi \in \mathcal{C}$ there exists $\epsilon>0$ and a unique maximal solution ${\bf z}: [0,\epsilon] \rightarrow \mathds{R}^n$ to $\dot {\bf  z}(t)=f({\bf z}(t))$  satisfying ${\bf z}(0)=\xi$ and ${\bf z}(t)\in \mathcal{C}$ for all $t\in [0,\epsilon]$.
%\end{enumerate}
The implementations of a hybrid control system differ in the sequence of elections of jumping and flowing. Therefore, we can think about two particular implementations obtained by simply always choosing to jump ({\em jumping-first} solution) or always choosing to flow ({\em flowing-first} solution). Following this idea, for $\mathcal{H} = (\mathcal{C}, f , \mathcal{D}, g)$, let us define the following two hybrid systems: 
\begin{equation}
\mathcal{H}^\mathcal{D} = (\mathcal{C}\setminus \mathcal{D}, f, \mathcal{D}, g)
\end{equation}
\begin{equation}
\mathcal{H}^\mathcal{C} = (\mathcal{C}, f, \mathcal{D}\setminus \mathcal{C}^*, g),
\end{equation}
where
\begin{equation}
\mathcal{C}^* =\{{\bf x} \in \mathcal{C}:T_\mathcal{C}(\mathbf{x})\cap f(\mathbf{x})\neq \varnothing \},
\end{equation}
being $T_\mathcal{C}(\mathbf{x})$ the tangent cone\footnote{The tangent cone to the set $\Sigma \subset \mathds{R}^n$ at $\mathbf{x} \in \mathds{R}^n$, $T_\Sigma(\mathbf{x})$,  is the set of all vectors $\mathbf{w}\in \mathds{R}^n$ for which there exist $\mathbf{x}_i \in \Sigma$, $\tau_i >0$, for all $i=1,2,...$ such that $\mathbf{x}_i \rightarrow \mathbf{x}$, $\tau_i \rightarrow 0$, and $(\mathbf{x}_i -\mathbf{x})/\tau_i \rightarrow \mathbf{w}$ as $i \rightarrow \infty$. Informally speaking, $\mathcal{C}^*$ is basically the set of points in $\mathcal{C}$ from which flowing to $\mathcal{C}$ is possible.} to $\mathcal{C}$ at the point ${\bf x}$. 

Finally, note that $f$ and $g$ are not set-valued functions, and thus, considering the hybrid basic conditions and the basic uniqueness conditions (see App. A and B), it directly follows the following result on the existence of implementations. 

\vspace{0.15cm}
{\bf Corollary 3.2} {\em Consider a hybrid control system $\mathcal{H}$, satisfying the basic hybrid conditions, and that for every $\xi \in \mathcal{C}$ there exists $\epsilon>0$ and a unique maximal solution ${\bf z}: [0,\epsilon] \rightarrow \mathds{R}^n$ to $\dot {\bf  z}(t)=f({\bf z}(t))$  satisfying ${\bf z}(0)=\xi$ and ${\bf z}(t)\in \mathcal{C}$ for all $t\in [0,\epsilon]$. The hybrid systems $\mathcal{H}^\mathcal{C}$ and $\mathcal{H}^\mathcal{D}$ are  implementations of $\mathcal{H}$.}

\vspace{0.15cm}
Thereafter, we refer to $\mathcal{H}^\mathcal{C}$  and $\mathcal{H}^\mathcal{D}$ as {\em flowing-first implementation} and {\em jumping-first implementation}, respectively. The definition of $\mathcal{H}^\mathcal{C}$ is a bit more involved, since a simple definition of the jump set, as $\mathcal{D}_I=\mathcal{D} \setminus \mathcal{C}$, may lead to the existence of maximal solutions to $\mathcal{H}^\mathcal{C}$ that are not maximal solutions to $\mathcal{H}$. In order to build the jump set of the implementation it is necessary to add to $\mathcal{D}_I$ all the points for which flowing is not possible. 

Note that when $\mathcal{H}$ has unique solutions then there is no possibility of jumping/flowing choice. In this case $\mathcal{H}^\mathcal{C} = \mathcal{H}^\mathcal{D} = \mathcal{H}$, in the sense that the three hybrid systems produce the same unique solution for each initial point. 

\subsection{A simple example}
\begin{figure}[t] %  figure placement: here, top, bottom, or page
  % \small
   \centering
     \subfigure[\label{fig:flowing-first}]{\includegraphics[width=4cm]{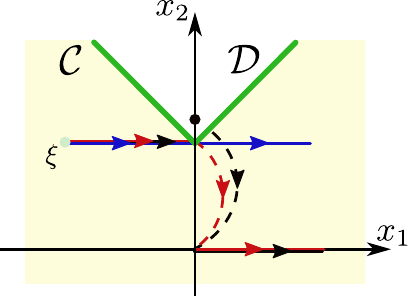}}
   \subfigure[\label{fig:jumping-first}]{\includegraphics[width=4cm]{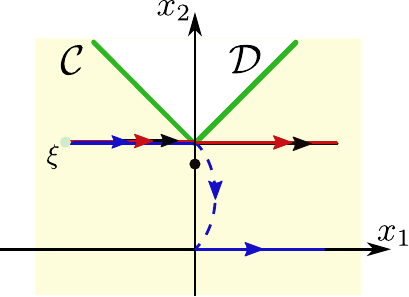}}
   \caption{Perturbation-free (blue) and perturbed (red) solutions to \eqref{eq-ex1H}% with jump set  
% $\mathcal{D} = \{ (x_1,x_2) \in \mathds{R}^2:x_1-x_2 \leq -1, -x_1-x_2 \leq -1
%\}
%$ and flow set $\mathcal{C} = \overline{\mathds{R}^2 \setminus \mathcal{D}}$
:  ({\em left}) Flowing-first solution $\phi^\mathcal{C}$ and perturbed solution $\phi_{\delta \mathbf{n}_{1a}}$, ({\em right})  Jumping-first solution $\phi^\mathcal{D}$ and perturbed solution $\phi_{\delta \mathbf{n}_{1b}}$. The perturbation signal is shown added to the perturbed solution (black).}
\label{fig-ex1}
\end{figure}

Although this example is not a hybrid control system, it is developed here to make clear the relation between hybrid  systems and its implementations, that will make full sense in the hybrid control system example of Section 3.3. Consider the hybrid system $\mathcal{H}$ on $\mathds{R}^2$ given by 
\begin{equation}
%\small
\mathcal{H}:  \left\{  \begin{array}{lcllcr}
\mathbf{\dot{x}} = \left( \begin{array}{c} 1 \\ 0 \end{array} \right), \quad  &\mathbf{x}  \in \mathcal{C}, \\
\mathbf{x}^+ =  \left( \begin{array}{c} 0 \\ 0 \end{array} \right), \quad &\mathbf{x} \in \mathcal{D}, %\\
%\mathbf{x}(0) &= \xi 
\end{array}
\right.
\label{eq-ex1H}
\end{equation}
where the jump set is the convex polytope $\mathcal{D} = \{ (x_1,x_2) \in \mathds{R}^2:x_1-x_2 \leq -1, -x_1-x_2 \leq -1
\}
$, and the flow set is $\mathcal{C} = \overline{\mathds{R}^2 \setminus \mathcal{D}}$ (see Fig. \ref{fig-ex1}). It is easy to see that all the maximal solutions to $\mathcal{H}$ are complete, and thus $\mathcal{H}$ is forward complete from any compact set $K \subset \mathds{R}^2$. In addition, since $\mathcal{H}$ satisfies the basic hybrid conditions (note that the flow and jump maps are constant, and the flow and jump sets are closed). \textcolor{black}{Thus, by Corollary 2.2, $\mathcal{H}$ is robust to perturbations for any compact set $K \subset \mathds{R}^2$. Next, we analyze the robustness of its implementations for the set $K = \{\xi\} = \{(-1,1)\}$}.

\vspace{0.15cm}
{\em Perturbation-free solutions}. It directly follows that there are only two solutions for the initial point $\xi$, which are $\phi^\mathcal{C}:[0,\infty)\times \{0\} \rightarrow \mathds{R}^2$ with $\phi^\mathcal{C}(t,0)=(-1+t,1)$,  $t\in[0,\infty)$ and $\phi^\mathcal{D}:[0,1]\times \{0\} \cup [1,\infty)\times \{1\} \rightarrow \mathds{R}^2$ with $\phi^\mathcal{D}(t,0)=(-1+t,1)$ for all $t\in[0,1]$ and $\phi^\mathcal{D}(t,1) = (-1+t,0)$ for all $t\in[1,\infty)$. The solutions $\phi^\mathcal{C}$ and $\phi^\mathcal{D}$ are plotted in Fig. \ref{fig:flowing-first} and \ref{fig:jumping-first}, respectively. 

\vspace{0.15cm}
{\em Perturbed solutions}. \textcolor{black}{The perturbed hybrid system is $\mathcal{H}_{\delta(\mathbf{n}_1,\mathbf{n}_2,\mathbf{n}_3)}$, where by simplicity  $\mathbf{n}_2=\mathbf{n}_3 = 0$, that is only state perturbations are considered.
Consider the admissible perturbation signal $\mathbf{n}_{1a}:[0,1]\times \{0\} \cup [1,\infty)\times \{1\} \rightarrow \mathds{R}^2$, given by $\mathbf{n}_{1a}(1,0) = (0,1)$ and $\mathbf{n}_{1a}(t,j) = (0,0)$ otherwise.  For any $\delta >0$, the hybrid arc $\phi_{\delta \mathbf{n}_{1a}}:[0,1]\times \{0\} \cup [1,\infty)\times \{1\} \rightarrow \mathds{R}^2$ is the unique solution to $\mathcal{H}_{\delta (\mathbf{n}_{1a},0,0)}$ with $\phi_{\delta \mathbf{n}_{1a}}(0,0) = \xi$, $\phi_{\delta \mathbf{n}_{1a}}(t,0)=(-1+t,1)$ for $t\in[0,1]$, and $\phi_{\delta \mathbf{n}_{1a}}(t,1) = (-1+t,0)$ for $t\in[1,\infty)$. Now consider the admissible perturbation signal $\mathbf{n}_{1b}: [0,\infty)\times \{0\} \rightarrow \mathds{R}^2$, with $\mathbf{n}_{1b}(1,0) = (0,-1)$ and $\mathbf{n}_{1b}(t,0) = (0,0)$ if $t\neq 1$. For any $\delta >0$, the hybrid arc $\phi_{\delta \mathbf{n}_{1b}}:[0,\infty)\times \{0\} \rightarrow \mathds{R}^2$ is the unique solution to $\mathcal{H}_{\delta (\mathbf{n}_{1b},0,0)}$, with $\phi_{\delta \mathbf{n}_{1b}}(0,0) = \xi$, is $\phi_{\delta \mathbf{n}_{1b}}(t,0)=(-1+t,1)$ for $t\in[0,\infty)$. The solutions $\phi_{\delta \mathbf{n}_{1a}}$ and $\phi_{\delta \mathbf{n}_{1b}}$ are plotted in Fig. \ref{fig:flowing-first} and \ref{fig:jumping-first}, respectively. }

\vspace{0.15cm}
{\em Robustness analysis}. For the chosen $K = \{\xi\}$, that $\mathcal{H}$ is robust to perturbation means that for any perturbed solution there exists a close perturbation-free solution. For example, considering the solution $\phi_{\delta \mathbf{n}_{1b}}$, it directly follows that $\phi^\mathcal{C}$ is $(T,J,\epsilon)$-close for any $T$, $J$ and $\epsilon$. The same applies to $\phi_{\delta \mathbf{n}_{1a}}$ and $\phi^\mathcal{D}$. This is the exact meaning of robustness to perturbations in the HI framework. Now let us analyze the implementations. First, note that for any implementation $\mathcal{H}^I$, the hybrid arcs $\phi_{\delta \mathbf{n}_{1a}}$ and $\phi_{\delta \mathbf{n}_{1b}}$ are solutions to $\mathcal{H}^I_{\delta \mathbf{n}_{1a}}$ and $\mathcal{H}^I_{\delta \mathbf{n}_{1b}}$, respectively. In addition, for any implementation, one of the hybrid arcs $\phi^\mathcal{C}$ or $\phi^\mathcal{D}$ is solution to $\mathcal{H}^I$. Suppose that $\phi^\mathcal{D}$ is solution to an implementation $\mathcal{H}^I$, then for the implementation to be robust to perturbations, both solutions $ \phi_{\delta \mathbf{n}_{1a}}$ and $ \phi_{\delta \mathbf{n}_{1b}}$ should be $(T,J,\epsilon)$-close to $\phi^\mathcal{D}$ for a small enough $\delta$, since $\phi^\mathcal{D}$ is the unique solution. However, it is clear that $\phi^\mathcal{D}$ and $\phi_{\delta \mathbf{n}_{1b}}$ are not $(T,J,\epsilon)$-close  for $J=1$ independently of $\delta$ (see Fig 2.b). Similarly for $\phi^\mathcal{C}$ and $\phi_{\delta \mathbf{n}_{1a}}$(see Fig 2.a). As a result, there are not implementations of $\mathcal{H}$ that are robust to perturbations for the set $K$.

\subsection{A hybrid control system example}

The FORE ({first order reset element}) controller was introduced in \cite{Horowitz75}, and since then it has been used in a number of works  (see for example \cite{libroResetBB} and references therein). Different versions of FORE has been devised in the literature, some of them in the HI framework; here, the FORE proposed in \cite{Nesic11} is used: 
\begin{equation}
\left\{  \begin{array}{lll}
{\dot{{\tau}}}=1,  &  {\dot{{x}}_c}  = -\lambda x_r + v , \quad   & \epsilon v^2+2x_r v   \geq 0 \text{ or } \tau \leq \rho,\ \\
{{{\tau}^+}}=0,  & {{x}}_c^+ = 0, \quad &  \epsilon v^2+2x_r v   \leq 0 \text{ and } \tau \geq \rho,
\end{array}
\right. 
\label{eq-ex2H}
\end{equation}
where $(x_r,\tau) \in \mathds{R}\times \mathds{R}_{\geq 0}$ is the state, $x_r$ is the output, and $v \in \mathds{R}$ is the input.  %the flow and jump sets are defined as ${\mathcal C} = \{(\tau,x_r) \in \mathds{R}_{\geq 0}\times \mathds{R} : \epsilon v^2+2x_r v   \geq 0 \text{ or } \tau \leq \rho\}$, and ${\mathcal D} = \{(\tau,x_r) \in \mathds{R}_{\geq 0}\times \mathds{R} : \epsilon v^2+2x_r v   \leq 0, \tau \geq \rho\}$, respectively. 
In addition, $\lambda \in \mathds{R}$ defines the pole of the base system, and $\rho$ and $\epsilon$ are some positive constants. See \cite{Nesic11}-Section III for details and motivation.

Now, consider a hybrid control system $\mathcal{H}^\text{cl}$ (Fig. 1) consisting of the feedback interconnection of a plant $P$ with transfer function $P(s) = \frac{s+1}{s(s+0.2)}$ and a FORE. This feedback control system has been analyzed in a number of works, including several works in the HI framework (\cite{Nesic05,Zaccarian11}). 
 
If the input and state of $P$ are $u$ and $(x_1,x_2)$, respectively, then the feedback interconnection is given by $u = x_r$ and $v = - x_2$. The closed-loop hybrid system $\mathcal{H}^\text{cl}$, with state $(\mathbf{x},\tau) = (x_1, x_2, x_r,\tau)\in \mathds{R}^3 \times \mathds{R}_{\geq 0}$, %and output $y = x_2$, 
is given by
\begin{equation}
%\small
\mathcal{H}^\text{cl}: 
 \left\{  \begin{array}{lll}
{\dot{\mathbf{\tau}}}= 1, & {\dot{\mathbf{x}}}  = A\mathbf{x}=\left( \begin{array}{ccc} 0&0&1 \\ 1 &-0.2&1\\0&-1&-1 \end{array} \right)\mathbf{x} ,&(\mathbf{x},\tau) \in {\mathcal C},\\
{\mathbf{\tau}}^+=0, & {\mathbf{x}}^+ = A_R\mathbf{x} = \left( \begin{array}{ccc} 1&0&0 \\ 0 &1&0\\0&0&0 \end{array} \right)\mathbf{x} , & (\mathbf{x},\tau) \in {\mathcal D}, %\\
%\mathbf{x}(0) &= \xi 
\end{array}
\right.
\label{eq-ex2Hcl}
\end{equation}
%\epsilon x_2^2+2x_2x_r  \leq 0 
%\epsilon x_2^2+2x_2x_r \geq 0
where $\lambda =1$ has been chosen. Here, the flow and jump sets are given by ${\mathcal C} = \{(\mathbf{x},\tau)\in \mathds{R}^3 \times \mathds{R}_{\geq 0} : \epsilon x_2^2-2x_2x_r  \geq 0  \text{ or } \tau \leq \rho\}$, and ${\mathcal D} = \{(\mathbf{x},\tau) \in \mathds{R}^3 \times \mathds{R}_{\geq 0} : \epsilon x_2^2-2x_2x_r \leq 0 \text{ and } \tau \geq \rho\}$, respectively.  Note that if $(\mathbf{x},\tau) \in \mathcal{D}$ then $(\mathbf{x}^+,\tau^+) = (x_1,x_2,0,0) \in \mathcal{C}\setminus  \mathcal{D}$, and thus only flowing is possible after a jump. 

Although ${\mathcal H}$ is defined on $\mathds{R}^3 \times \mathds{R}_{\geq 0}$, $\tau$ is a controller state component that acts simply as a timer to avoid that two consecutive jumps are performed in lesser time than the minimum dwell time $\rho$. Note that $\mathcal{H}$ satisfies the hybrid basic conditions and is forward complete from any compact set $K \subset \mathds{R}^3 \times \{ 0\}$; thus, Corollary 2.2 guarantees that $\mathcal{H}$ is robust to perturbations for any compact set $K$. 
\begin{figure}[t] %  figure placement: here, top, bottom, or page
   %\small
   \centering
   \subfigure{\includegraphics[width=6.25cm]{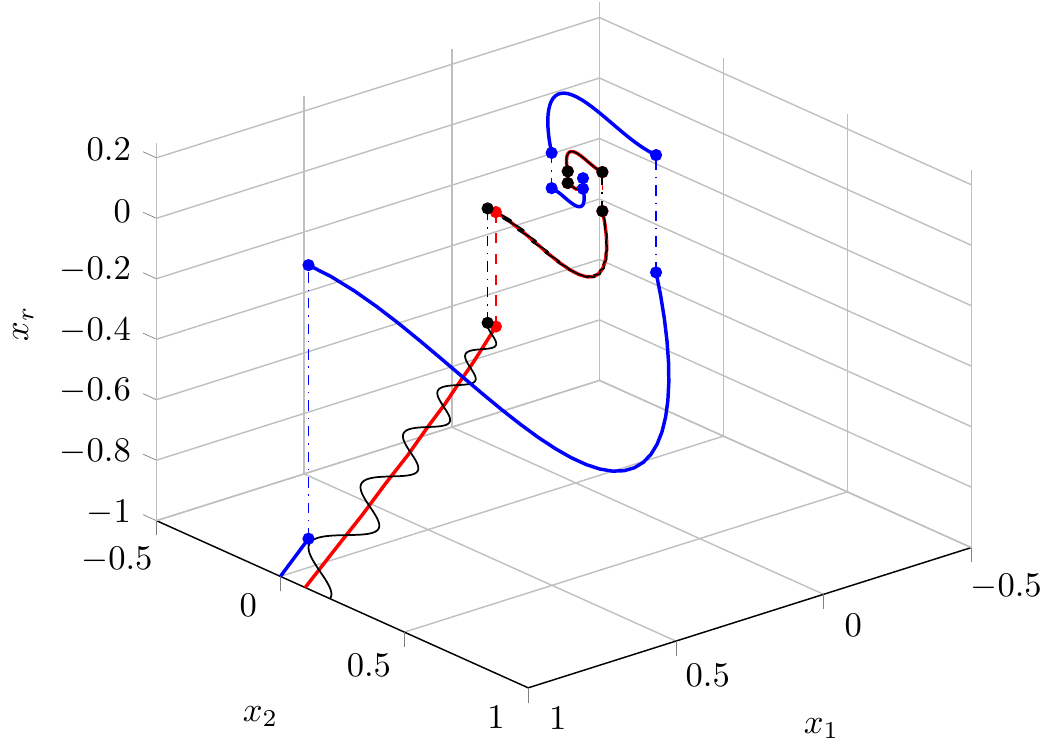}}
   \subfigure{\includegraphics[width=6.25cm]{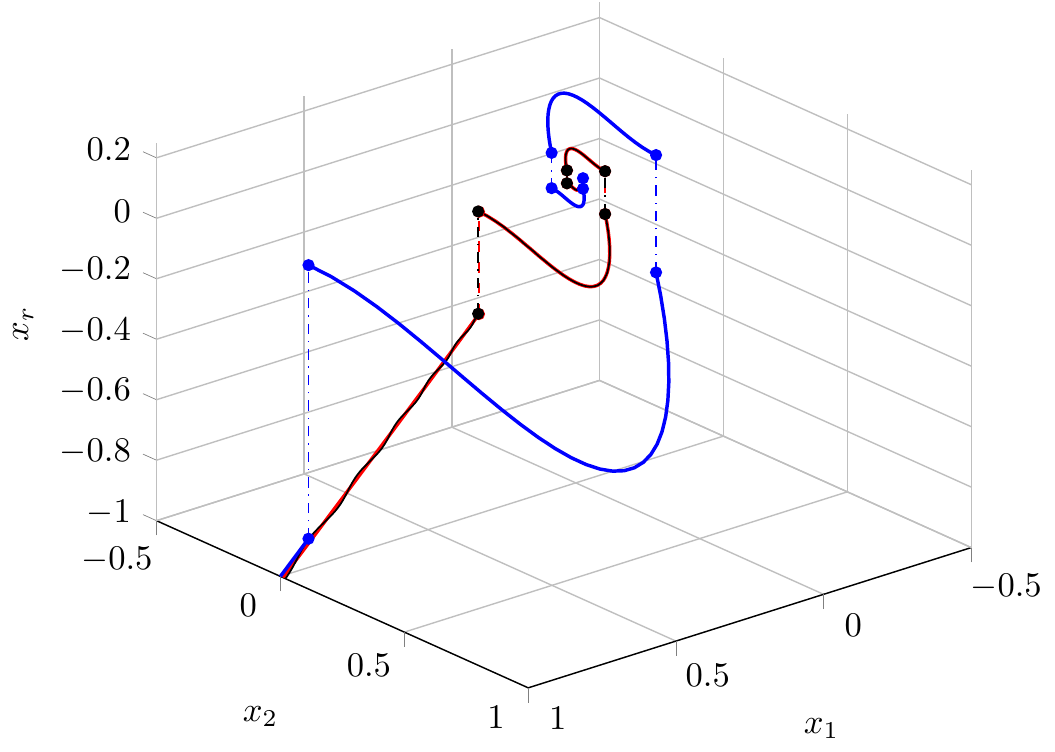}}
   \subfigure{\includegraphics[width=6.25cm]{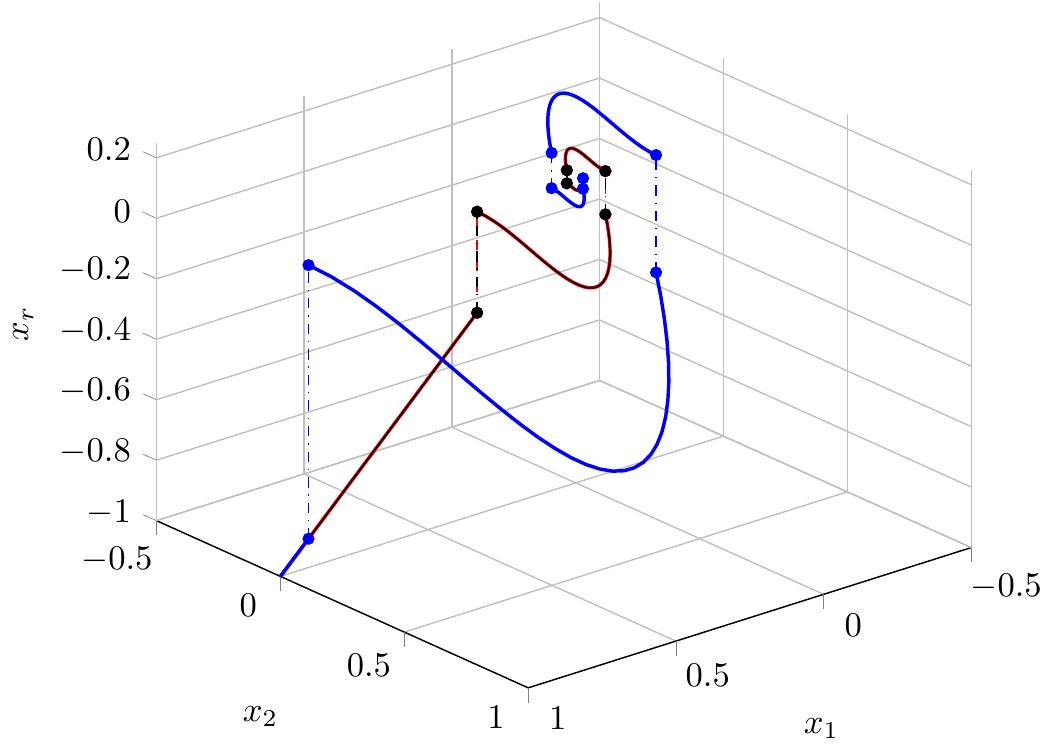}}
   \caption{Noise-free solution $\phi^{\mathcal D}$ (blue) with $\phi^{\mathcal D}(0,0) = (1,0,-1,0)$,  and noisy solutions $\phi_{\delta {n}_a}$ (red),  with $\phi_{\delta {n}_a}(0,0) = (1,\delta,-1,0)$ and ${n}_a(t)= e^{-t}\cos(10\pi t)$: (top) $\delta = 0.1$, (middle) $\delta = 0.01$, (down) $\delta = 10^{-6}$. The perturbation signal $\mathbf{e}_1 = (0,\delta n_a,0,0)$ (black) is shown added to the noisy solution .
   }  
   \label{fig-ex2D}
\end{figure}
In this example, we only consider the measurement noise $\mathbf{d}_1$ as the unique perturbation affecting the state $x_2$ in the feedback path, that is $\mathbf{d}_1 = (e,0)$ and $\mathbf{d}_2 = (0,0)$, for some scalar perturbation signal $e$, and thus the perturbed hybrid control system is $\mathcal{H}^\text{cl}_{(\mathbf{d_1},\mathbf{0})}$ (see Fig. 1). 

As a result, the perturbed control system takes the form (7), where by using (6) it results that $\mathbf{e}_1 = (0,e,0,0)$, $\mathbf{e}_2 = (0,0.2e,0,0)$, and $\mathbf{e}_3 = (0,-e,0,0)$, and thus the admissible perturbation signals are $\mathbf{n}_1 = (0,n,0,0)$, $\mathbf{n}_2 = (0,0.2n,0,0)$, and $\mathbf{n}_3 = (0,-n,0,0)$, for some scalar admissible perturbation signal $n$. In the following, different noise-free and noisy solutions to the hybrid control system are analyzed. Two admissible perturbation signals $n$ will be used: ${n}_a(t)= e^{-t} \cos(10\pi t)$ and ${n}_b(t)=\cos(10\pi t)$. For simplicity, the notation $\mathcal{H}^\text{cl}_{\delta {n}_a}$ or $\mathcal{H}^\text{cl}_{\delta {n}_b}$ will be used for the perturbed hybrid control systems, respectively. On the other hand, 
 $K$ will be any compact subset of $\mathds{R}^3 \times \{ 0\}$ such that $(\xi,0) \in K$, where $\xi = (1,0,-1)$. In addition, the values $\epsilon = 0.1$ and $\rho = 0.1$ have been chosen.

\vspace{0.1cm}

\begin{figure}[t] %  figure placement: here, top, bottom, or page
   \small
   \centering
   \subfigure{\includegraphics[width=6.25cm]{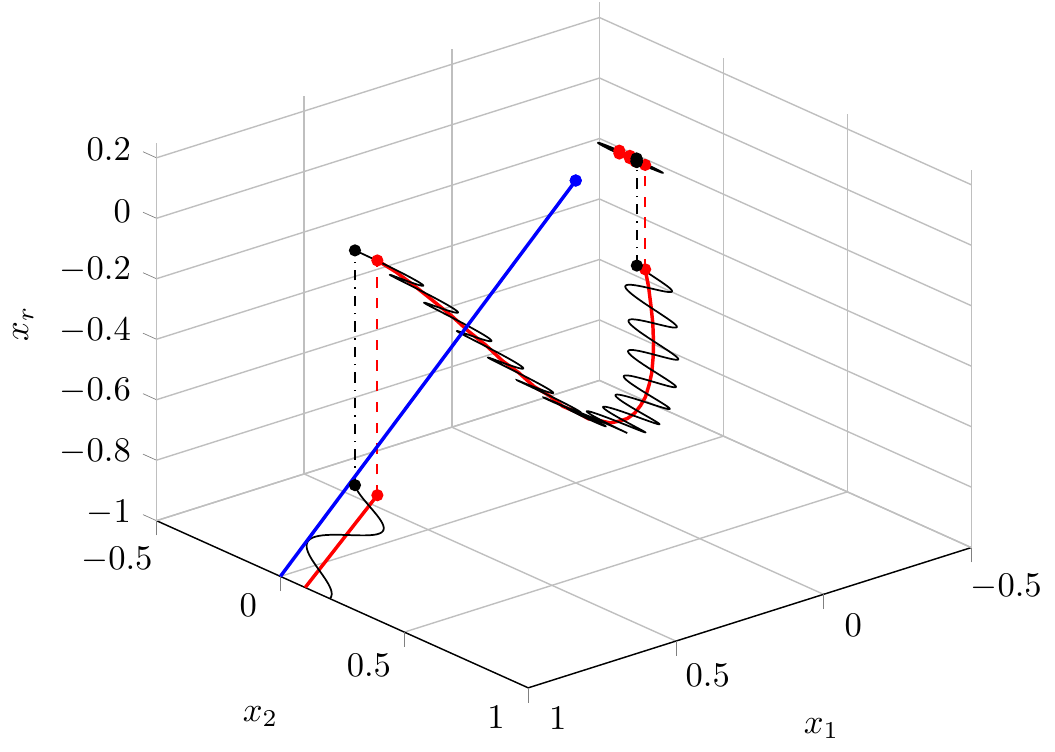}}
   \subfigure{\includegraphics[width=6.25cm]{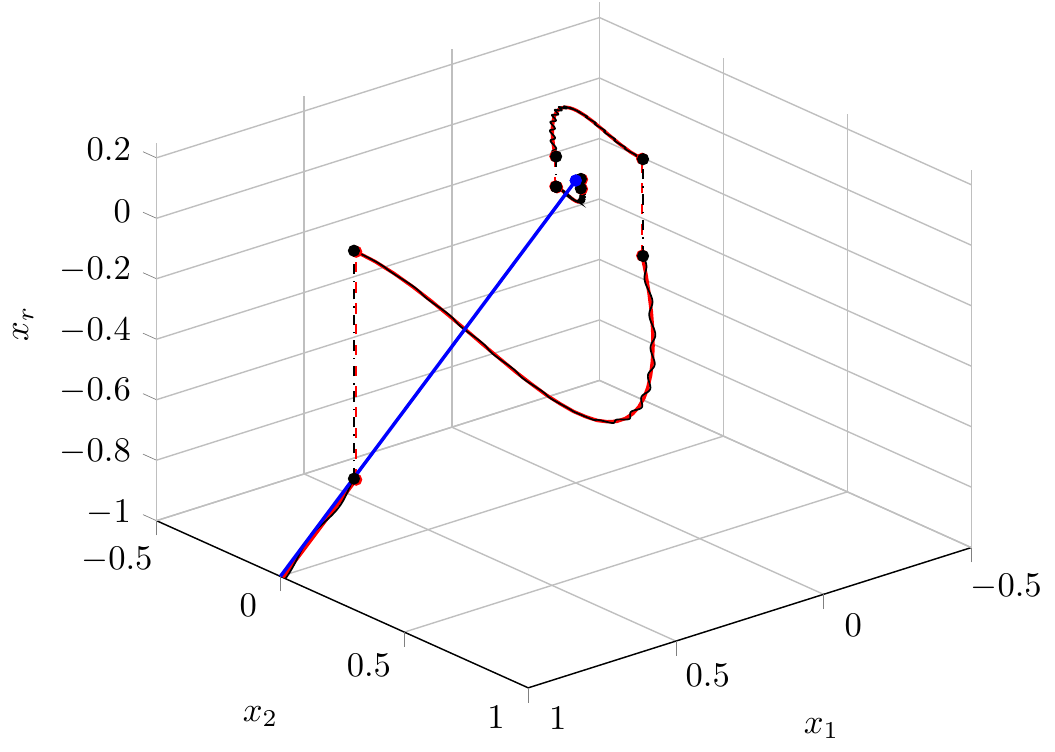}}
   \subfigure{\includegraphics[width=6.25cm]{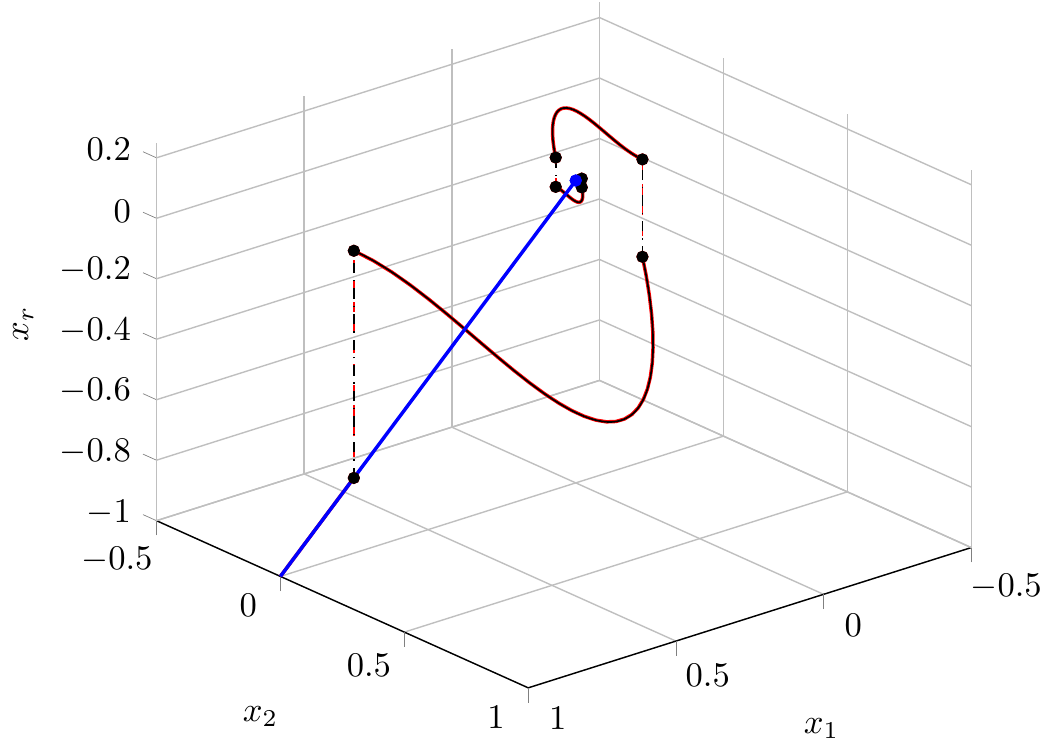}}
   \caption{Noise-free solution $\phi^{\mathcal C}$ (blue)  with $\phi^{\mathcal C}(0,0) = (0,1,0,-1)$ and noisy solutions $\phi_{\delta{n}_b}^{\mathcal C}$ (red), with $\phi_{\delta {n}_b}^{\mathcal C}(0,0) =  (0,1,\delta,-1)$, and ${n}_b(t)= \cos(10\pi t)$: (top) $\delta = 0.1$, (middle) $\delta = 0.01$, (down) $\delta = 10^{-6}$. The perturbation signal $\mathbf{e}_1 = (0,\delta n_b,0,0)$ (black) is shown added to the noisy solution.
   }  
   \label{fig-ex2C}
\end{figure}

%and thus the admissible noise $\mathbf{n}$ is given by $\mathbf{n} = (0,0,n,0)$ for some signal $n:\mathds{R}_{\geq 0} \rightarrow \mathds{R}$. Consider $\xi = (1,0,-1)$ and some compact set $K$ such that $(0,\xi) \in K$; and also the noise signals ${n}_1(t)= e^{-t} \cos(10\pi t)$ and ${n}_2(t)=\cos(10\pi t)$. In addition, the values $\epsilon = 0.1$ and $\rho = 0.1$ has been chosen. Next, we analyze the robustness to noise for the implementations of $\mathcal{H}$ for the set $K$.\\
{\em Noise-free solutions}.  Any solution $\phi$ to $\mathcal{H}^\text{cl}$, with $\phi(0,0) = (\xi,0)$, has either a domain given by $\text{dom } \phi = [0 ,t_1]\times \{0\} \cup [t_1 ,t_2]\times \{1\} \cup \cdots$ with $t_1\geq \rho$, or a domain $\text{dom } \phi = [0 ,\infty) \times \{0\}$. By Def. 3.1, for any implementation $\mathcal{H}^{\text{cl},I}$, there is a solution $\phi$  to $\mathcal{H}^\text{cl}$ with one of the above domains, that is also solution to $\mathcal{H}^{\text{cl},I}$. By convenience, define the implementations set $\mathcal{H}^{\text{cl},I}_{t_1}$ with parameter ${t_1}\geq \rho$, as the set of all implementations for which the solution $\phi$ has the first jump at $t_1$ (if the domain of the solution is $\text{dom } \phi = \{[0 ,\infty),0\}$ then the set is $\mathcal{H}^{\text{cl},I}_\infty$). On the other hand, solutions $\phi^\mathcal{D}$ and $\phi^\mathcal{C} $ of the jumping-first implementation $\mathcal{H}^{\text{cl},\mathcal{D}} \in \mathcal{H}^{\text{cl},I}_{\rho}$ and the flowing-first implementacion $\mathcal{H}^{\text{cl},\mathcal{C}} \in \mathcal{H}^{\text{cl},I}_\infty$ are plotted in Fig. \ref{fig-ex2D} and \ref{fig-ex2C} (simulations have been performed using \cite{toolbox}), respectively.
\vspace{0.1cm}

{\em Noisy solutions}. First, let us focus on the solutions $\phi_{\delta {n}_a}$ to $\mathcal{H}^\text{cl}_{\delta {n}_a}$. It is not difficult to see (details are omitted by brevity) that for any solution $\phi_{\delta {n}_a}$ with $\phi_{\delta {n}_a}(0,0)=(\xi,0)+(0,\delta,0,0)$, the domain is $\text{dom }\phi_{\delta {n}_a}= [0 ,s_1] \times\{0\} \cup [s_1 ,s_2]\times \{1\} \cup \cdots$ with $s_1\approx1.0977$, independently of $\delta$. On the other hand, for any solution $\phi_{\delta {n}_b}$ to $\mathcal{H}^\text{cl}_{\delta {n}_b}$ with $\phi_{\delta {n}_b}(0,0)=(\xi,0)+(0,\delta,0,0)$, the domain is $\text{dom }\phi_{\delta {n}_b}= [0 ,l_1]\times \{0\} \cup [l_1 ,l_2]\times \{1\} \cup \cdots$ with $l_1\approx 1.4430$, independently of $\delta$. Several noisy solutions are plotted in  Fig. \ref{fig-ex2D} and \ref{fig-ex2C} for different values of $\delta$.

\vspace{0.1cm}

{\em Robustness analysis}. Consider any $t_1 \geq \rho$, and any implementation $\mathcal{H}^{\text{cl},I} \in \mathcal{H}^{\text{cl},I}_{t_1}$. First, note that the truncation $\bar \phi_{\delta {n}_a}$ of $\phi_{\delta {n}_a}$ with $\text{dom } \bar \phi_{\delta {n}_a}= \{ (t,j)\in \text{dom } \phi_{\delta {n}_a}: t\leq s_1, \ j=0\}$ is a truncation of the solution  $\phi^I_{\delta {n}_a}$ to  $\mathcal{H}^I_{\delta {n}_a}$. Similarly for the solution $\phi^I_{\delta {n}_b}$ to $\mathcal{H}^I_{\delta {n}_b}$. Since the solutions of $\mathcal{H}^I$ are unique, for the implementation $\mathcal{H}^I$ to be robust to perturbations and for the set $K$, the solution $\phi$ to $\mathcal{H}^I$ with $\phi(0,0)=(\xi,0)$ must be $(T,J,\epsilon)$-close to $\phi^I_{\delta {n}_a}$ and $\phi^I_{\delta {n}_b}$ for any $T$, $J$, and $\epsilon$. However, considering $T\geq t_1$ and $J=0$, it is deduced from the truncations $\bar \phi_{\delta {n}_a}$  and $\bar \phi_{\delta {n}_b}$ that it is always possible to find a small enough $\epsilon$ such that $\max(|t_1-s_1|, |t_1-l_1|) >\epsilon$ (note that $s_1\neq l_1$), and thus, $\phi$ is not $(T,J,\epsilon)$-close to $\phi^I_{\delta {n}_a}$ or  $\phi^I_{\delta {n}_b}$. This means that any implementation in the sets $\mathcal{H}^I_{t_1}$, for $t_1 \geq \rho$, as hybrid system by themselves (for example the implementations $\mathcal{H}^\mathcal{D}$ and $\mathcal{H}^\mathcal{C}$), fail to satisfy the robustness to perturbations property, in spite of the fact that $\mathcal{H}$ satisfies that property. In contrast to the example of Section 3.2, in this case this is due to the existence of a subspace of $\mathds{R}^3$ that is invariant with respect to the flowing dynamic of the state ${\bf x}$; this is the unobservable subspace given by $\text{span}\{(1,0,-1)\}$.   
%Therefore, it is true that $\text{span}\{(1,0,-1)\} \subset \mathcal{C} \cap \mathcal{D}$.

\section{A new definition of robustness to perturbations}
Although hybrid basic conditions are sufficient for a hybrid control system to be robust to perturbations (according to Def. 2.1), this sense of robustness is not enough in control practice. It has been shown that implementations of a robust hybrid control system are not necessarily robust to perturbations.
% that measurement errors, in particular measurement noise, can force a rich behavior of solutions to hybrid control systems that may diverge of the noise-free solutions. %Besides tangential crossing or more generally grazing, more subtle behavior (for example due to existence of unobservable and invariant subspaces for the flow dynamics) may be possible in general.
%Although hybrid basic conditions are sufficient for a hybrid control system to be robust (according to Definition II.1), this sense of robustness is not enough in control practice. This is due to the fact that implementations of a hybrid control system are not necessarily robust to measurement errors. %, which is basic limitation of the HI approach. 
To overcome this limitation, a narrower notion of robustness to perturbations, that will be useful to characterize robustness of implementations, is proposed. %, based on the closeness of the noise-free solutions to noisy solutions (in a inner sense, in addition to the outer sense previously developed as given in Definition II.1). 
 In addition, a relationship with previously developed relaxations results for hybrid systems is developed.

\subsection{Strong Robustness to perturbations}

{\bf Definition 4.1} (Strong robustness to perturbations) {\em For a compact set $K \subset \mathds{R}^n$ such that the hybrid system $\mathcal{H}$ is forward complete from $K$, $\mathcal{H}$ is strongly robust to perturbations if it is robust to perturbations and, in addition, for any $\epsilon > 0$ and $(T, J) \in \mathds{R}_{\geq 0} \times \mathds{N}$ there exists $\delta^\ast > 0$ with the following property: for any admissible noise signal ${\mathbf n}$, any $\delta \in (0,\delta^\ast]$, any  $\xi \in K$, any $\xi_\delta \in \xi + \delta \mathds{B}$, and any solution $x \in \mathcal{S}_{\mathcal{H}}(\xi)$, there exists a solution $\mathbf{x}_{\delta \mathbf{n}}$ to $\mathcal{H}_{\delta \mathbf{n}}$, with $\mathbf{x}_{\delta \mathbf{n}}(0,0) = \xi_\delta$, such that $\mathbf{x}$ and $\mathbf{x}_{\delta \mathbf{n}}$ are $(T,J,\epsilon)$-close.}

\vspace{0.15cm}

Note that for any implementation, the properties of robustness and strong robustness to perturbations are equivalent, since they have unique solutions for any initial point.
%As it has been shown in Section III, the property of robustness to measurement errors of a hybrid system does not guaranty that there exist implementations of it which are also robust. 
Next, we show that a sufficient condition for the jumping-first and flowing-first implementations of a hybrid control system to be robust is that the hybrid control system be strongly robust.

%{\bf Corollary IV.2} {\em If the hybrid system $\mathcal{H}$ given by (1) is strongly robust to measurement noise, for some compact set $K \subset \mathds{R}^2$ such that $\mathcal{H}$ is forward complete from $K$, then for any solution $\mathbf{x} \in   \mathcal{S}_{\mathcal{H}}(K)$ there always exists an implementation $\mathcal{H}^I$, with $\mathbf{x} \in \mathcal{S}_{\mathcal{H}^I}(K)$, which is also strongly robust to measurement noise for the set $K$.}
\vspace{0.15cm}

{\bf Proposition 4.2} {\em Suppose that the hybrid control system $\mathcal{H}$, satisfying the assumptions of Corollary 3.2, % the hybrid basic conditions,
is strongly robust to perturbations for some compact set $K \subset \mathds{R}^n$, and in addition, $\mathcal{H}$ is forward complete from $K$. Then the flowing-first implementation, $\mathcal{H}^\mathcal{C}$, and the jumping-first implementation, $\mathcal{H}^\mathcal{D}$, are robust to perturbations for the set $K$.}

{\bf Proof:} Note that assumptions of Corollary 3.2 guarantee the existence of implementations $\mathcal{H}^\mathcal{C}$ and $\mathcal{H}^\mathcal{D}$. Let us prove that $\mathcal{H}^\mathcal{C}$ is robust to perturbations for the set K. A similar approach can be applied to $\mathcal{H}^\mathcal{D}$. Consider any $\epsilon>0$, any $(T, J) \in \mathds{R}_{\geq 0} \times \mathds{N}$, any admissible perturbations signal ${\bf n}$, any $\xi \in K$ and the unique solution $x \in \mathcal{S}_{\mathcal{H}^\mathcal{C}}(\xi)$, then we aim at finding $\delta^*>0$ in Def. 4.1, which may depend on $\epsilon$,  $T$, and $J$. From the definition of implementation (Def. 3.1), we get $x \in \mathcal{S}_{\mathcal{H}}(\xi)$. Since $\mathcal{H}$ is strongly robust to perturbations for the set $K$ then there exists $\bar \delta^*$ such that for any $\delta \in (0,\bar \delta^*]$, any $\xi_\delta \in \xi + \delta \mathds{B}$, there exists a solution $\mathbf{x}_{\delta \mathbf{n}}$ to $\mathcal{H}_{\delta \mathbf{n}}$, with $\mathbf{x}_{\delta \mathbf{n}}(0,0) = \xi_\delta$, such that $x$ and $\mathbf{x}_{\delta \mathbf{n}}$ are $(T,J,\epsilon)$-close.  Consider the truncation $\bar{\mathbf{x}}_{\delta \mathbf{n}}$ of $\mathbf{x}_{\delta \mathbf{n}}$ with $\text{dom } \bar{\mathbf{x}}_{\delta \mathbf{n}} = \{ (t,j)\in \text{dom }\mathbf{x}_{\delta \mathbf{n}}: t\leq T, \ j\leq J\}$, if the truncation is also a solution to $\mathcal{H}^C_{\delta \mathbf{n}}$ then it directly follows that $\mathcal{H}^\mathcal{C}$ is strongly robust perturbations for the set $K$ by taking $\delta^*=\bar \delta^*$. 

By way of contradiction, suppose that any $\mathbf{x}_{\delta \mathbf{n}} \in \mathcal{S}_{\mathcal{H}_{\delta \mathbf{n}}} (\xi)$ that is $(T,J,\epsilon)$-close to $\mathbf{x}$, its truncation $\bar{\mathbf{x}}_{\delta \mathbf{n}}$ with $\text{dom } \bar{\mathbf{x}}_{\delta \mathbf{n}} = \{ (t,j)\in \text{dom }\mathbf{x}_{\delta \mathbf{n}}: t\leq T, \ j\leq J\}$  is not a solution to $\mathcal{H}^C_{\delta \mathbf{n}}$. Then for any  $\bar{\mathbf{x}}_{\delta \mathbf{n}}$, there exists $(t,j) \in \text{dom }  \bar{\mathbf{x}}_{\delta \mathbf{n}}$ with $(t,j+1) \in \text{dom } \bar{\mathbf{x}}_{\delta \mathbf{n}}$ such that $\bar{\mathbf{x}}_{\delta \mathbf{n}}(t,j)+ \delta \mathbf{n}(t,j) \in \mathcal{D}\cap \mathcal{C}^*$. 

From the strong robustness  of $\mathcal{H}$, for any of the previous $(t,j)$ there exist $\bar \epsilon>0$ depending on $\epsilon$ with $\lim_{\epsilon\rightarrow0}\bar \epsilon =0$ and $s$ such that $|t-s|<\bar \epsilon$, $(s,j) \in \text{dom }\mathbf{x} $, $(s,j+1) \in \text{dom }\mathbf{x}$, and $ \| \mathbf{x}_{\delta \mathbf{n}}(t,j)- \mathbf{x}(s,j) \|< \bar \epsilon$. Therefore, we get
\begin{equation}
\nonumber
\begin{split}
\bar \epsilon+\delta &>\| \mathbf{x}_{\delta \mathbf{n}}(t,j) - \mathbf{x}(s,j)\| + \delta \geq \\  \| \mathbf{x}_{\delta \mathbf{n}}(t,j)+ \delta \mathbf{n}(t,j) - \mathbf{x}(s,j)\| 
&\geq \inf_{y\in  \mathcal{D} \cap  \mathcal{C}^*} \|y- \mathbf{x}(s,j) \|
\end{split}
\end{equation}
Since $\mathbf{x}(s,j) \in \mathcal{D} \setminus \mathcal{C}^*$ and $s< T+\bar \epsilon$  there exists $\gamma(T,\bar \epsilon)$ such that 
\begin{equation}
\nonumber
\bar \epsilon +\delta>\inf_{y\in \mathcal{D} \cap \mathcal{C}^*} \|y- \mathbf{x}(s,j) \| \geq \gamma(T,\bar \epsilon)
\end{equation}
Note that $\gamma$ is nonincreasing in $\bar \epsilon$. The above inequality must hold for any $\bar \epsilon$ and $\delta \in (0,\bar \delta^*)$. For a sufficiently small $\epsilon$ and $\delta$, it follows that $\bar \epsilon + \delta <\gamma(T,1) \leq \gamma(T,\bar \epsilon)$, which is a contradiction. Therefore, the truncation $\bar{\mathbf{x}}_{\delta \mathbf{n}}$ is solution to $\mathcal{H}^\mathcal{C}$, and the proof is complete.
$\Box$

\vspace{0.15cm}

A direct application of the strong robustness definition to Example 3.2 results in that $\mathcal{H}$, given by (12), is strongly robust to perturbations for any compact set $K\subset \mathds{R}^2 \setminus \mathcal{X}$, where $\mathcal{X}=\{(x_1,x_2)\in \mathds{R}^2:x_1 \geq 0,x_2 = x_1 +1\} \cup \{(x_1,x_2) \in \mathds{R}^2 : x_1 \leq 0, x_2 = 1\}$. Moreover, both implementations $\mathcal{H}^\mathcal{D}$ and $\mathcal{H}^\mathcal{C}$ are robust for any compact set $K \subset \mathds{R}^2 \setminus \mathcal{X}$. Example 3.3 shows that besides avoiding grazing, the set $K$ cannot contain some specific initial points; in general, higher order hybrid control systems requires a deeper analysis. %This suggests that the proposed strong robustness notion may be hard to be developed for a large class of hybrid control systems, and thus it may be more convenient to search for specific conditions that be useful for particular classes of hybrid control systems. Following, a useful relationship with hybrid relaxation results is given.
In the following, a useful relationship with hybrid relaxation results is given,  providing a path for characterization of conditions that implies strong robustness.

\subsection{Relationship with hybrid relaxation results}
In \cite{Cai08}, several relaxation results are used to analyze continuous dependence on initial conditions of solutions to hybrid systems. Although the scope is more general than hybrid systems given by (1), it turns out that some of these relaxation results may be helpful to analyze strong robustness to perturbations of hybrid systems like (1). \textcolor{black}{A first result in that direction is the following proposition, that follows by using some relaxation properties (an extension of the strong relaxation property in \cite{Cai08}, see Appendix C). }
%Let us define the set of initial conditions $\mathcal{K}$  as $\mathcal{K} = (\mathcal{C}\setminus \mathcal{D})\cup(\mathcal{D}\setminus\mathcal{C})\cup (\text{int } \mathcal{C} \cap \text{int } \mathcal{D})$. 
%By simplicity, it will be assumed that the flow dynamics has a unique solution\footnote{For each possible flowing interval $[t_j,t_{j+1}]$, the initial value problem $\dot{\mathbf{x}}=f(\mathbf{x})$, $\mathbf{x}(t_j) = \mathbf{x}_j$ has a unique solution over $[t_j,t_{j+1}]$. A well-known sufficient condition is that $f$ be a Lipschitz function; note that this is the case of the Examples in Section 3.}. 

\vspace{0.15 cm}

%{\bf Proposition IV.3} {\em Consider a hybrid system $\mathcal{H}$ satisfying the hybrid basic conditions , and the total hybrid relaxation conditions\footnote{See Appendix C, the name is inspired in the classical concept of {\em total} stability for ordinary differential equations (also referred to as stability under persistent disturbances) \cite{Hahn67}.}, and a compact set $K\subset \mathcal{K}$ such that $\mathcal{H}$ is forward complete from $K$. Then $\mathcal{H}$ is strongly robust to measurement noise for the set $K$.
%}

{\bf Proposition 4.3} \textcolor{black}{{\em Consider a hybrid system $\mathcal{H}$ satisfying the hybrid basic conditions and a compact set $K\subset \mathds{R}^n$ such that $\mathcal{H}$ is forward complete from $K$. If for each $\xi \in K$ total strong relaxation is possible\footnote{See Appendix C, the name is inspired in the classical concept of {\em total} stability for ordinary differential equations (also referred to as stability under persistent disturbances) \cite{Hahn67}.} for solutions from $\xi$ then $\mathcal{H}$ is strongly robust to perturbations for the set $K$.
}}

{\bf Proof.} 
Since $\mathcal{H}$ satisfies the hybrid basic conditions, the robustness to perturbations is directly obtained by Corollary 2.2, and thus the proof is centered on the additional property for strong robustness according to Def. 4.1. 
In first place, using similar arguments to the proof of Th. 3.4 in \cite{Cai08}, it can be shown that total strong relaxation implies\footnote{This property may be referred to as that total strong relaxation for initially flowing (respectively, initially jumping) solutions from $\xi$ relative to $\mathcal{C}$ (respectively, relative to $\mathcal{D}$) is possible (using a direct analogy with Def. 3.1 in \cite{Cai08})} that given $\xi \in \mathds{R}^n$, for any compact solution $\mathbf{x}:\text{dom }\mathbf{x} \rightarrow \mathds{R}^n$ to $\mathcal{H}$ with $\mathbf{x}(0,0) = \xi$ and for any $\epsilon >0$, there exist $\delta >0$ such as for any admissible perturbation signal $\mathbf{n}$, any $\xi_\delta \in (\xi + \delta \mathds{B}) \cap (\mathcal{C} \cup \mathcal{D})$ there exist a solution $\mathbf{x}_{\delta \mathbf{n}}$ to the perturbed hybrid system $\mathcal{H}_{\delta \mathbf{n}}$ such as if $\mathbf{x}(T,J) \in \mathcal{D}$, where $(T,J) = \text{max dom }\mathbf{x}$, then $\mathbf{x}_{\delta \mathbf{n}}(\tau,J) \in \mathcal{D}$, where  $(\tau,J) = \text{max dom }\mathbf{x}_{\delta \mathbf{n}}$, and $\mathbf{x}$ and $\mathbf{x}_{\delta \mathbf{n}}$ are $(T,J,\epsilon)$-close. %The only difference with the inductive argument of the proof of Theorem 3.4 (\cite{Cai08}) is that instead of the argument that $g(\mathbf{x}_{\delta \mathbf{n}}(t'_J,J-1)) \in x(t_J,J) + \epsilon  \mathds{B}$, we use the fact that $g(\mathbf{x}_{\delta \mathbf{n}}(t'_J,J-1) + \delta \mathbf{n}(t'_J,J-1)) \in \mathbf{x}(t_J,J) + \epsilon \mathds{B}$ (which directly follows by continuity of $g$).

It remains to show uniformity with respect to the initial condition, that is that $\delta$ works for all solutions from $\xi \in \mathcal{K}$. By contradiction, if $\mathcal{H}$ is robust to perturbations but not strongly robust to perturbations then for some $\epsilon >0$, and $(T, J) \in \mathds{R}_{\geq 0} \times \mathds{N}$, there exist a sequence $\mathbf{x}_{i}:\text{dom }\mathbf{x}_{i} \rightarrow \mathds{R}^n$ of solutions to $\mathcal{H}$ with $\mathbf{x}_{i}= \xi$, a sequence of admissible perturbation signal $\mathbf{n}_i$, a sequence $\delta_i \rightarrow 0$, and a sequence $\xi_{\delta_i} \in  \xi+\delta_i \mathds{B}$, such as all solutions $\mathbf{x}_{{\delta_i}\mathbf{n}_i}$ to the hybrid system $\mathcal{H}_{{\delta_i}\mathbf{n}_i}$, with  $\mathbf{x}_{{\delta_i}\mathbf{n}_i}(0,0) = \xi_{\delta_i}$, satisfies that $\mathbf{x}_{i}$ and $\mathbf{x}_{{\delta_i}\mathbf{n}_i}$ are not $(T,J,\epsilon)$-close.  Similar arguments to proof of Proposition 6.2 in \cite{Cai08} may be applied, resulting in a contradiction of the total strong relaxation at $\xi$. Finally, the uniformity of $\delta$ in $\mathcal{K}$ comes of an argument similar to the one used in Corollary 6.4 in \cite{Cai08}, which ends the proof.
$\Box$

\vspace{0.15cm}

%See \cite{Cai08}, Section 6, for a detailed analysis of the related properties of strong relaxation and continuous dependence. Moreover,

 In \cite{Cai08}, some {\em hybrid relaxation conditions} are developed for strong relaxation for any $\xi \in  (\mathcal{C} \setminus \mathcal{D}) \cup (\mathcal{D} \setminus \mathcal{C}) \cup (\text{int } \mathcal{C} \cap \text{int } \mathcal{D}) $; it can be checked that these conditions are not satisfied for the Examples of Section 3, basically due to initial points that produce grazing in 3.2, and to the existence of a non-empty unobservable subspace in Section 3.3. This fact prevents the use of an extension of hybrid relaxation conditions to include perturbations, which would be of limited use in control practice. % could be performed, the limited application to our case 
 
 %As a result, in these cases {\em total} strong relaxation can not be implied by a possible extension of the hybrid relaxation conditions to cope with noise measurement and external disturbances.

Note that the difference between total strong relaxation and strong robustness is the uniformity in the latter, that is that $\delta$ works for all solutions from $\xi$ and for any $\xi \in K$, rather than for each solution we have a $\delta$; and thus, total strong relaxation is a property easier to check in principle. For example, for the hybrid control system of Section 3.2, it is not difficult to see that total strong relaxation is possible for solutions from any $\xi \in K$, for any compact $K \subset 
(\mathds{R}^3 \times \mathds{R}_{\geq 0})\setminus ((\mathcal{C} \cap \mathcal{D}) \cup (\text{span}\{(1,0,-1\}) \times \mathds{R}_{\geq 0}))$.

\section{Conclusions}
Robustness of hybrid systems to perturbations is a sound contribution of the HI framework, since it develops a property that may be applied to a generality of cases (hybrid systems satisfying the hybrid basic conditions). Although in general, this property of robustness is suitable for hybrid systems, hybrid control systems  demand a narrower property, since its implementations are not necessarily robust to perturbations, which is a clear limitation in control practice. This fact has been proved with two counterexamples (one specifically related to hybrid control systems),  showing that for two robust hybrid systems none of their implementations are robust. A new concept of robustness referred to as strong robustness to perturbations has been proposed; moreover, it has been shown that this new property is a sufficient condition for jumping-first and flowing-first implementations to be robust. Finally,  a relationship between strong robustness and previously developed hybrid relaxation results has been found.

%In order to avoid this problem, we propose a new concept of robustness to measurement noise, referred to as strong robustness, that is based on the closeness of the noise-free solutions to noisy solutions (in a inner sense, in addition to the outer sense previously developed in the HI framework). In addition, we provide sufficient conditions for the strong robustness to measurement noise based on some previously developed hybrid relaxation results.

%any perturbed solution (due to measurement noise and/or actuator disturbance) is arbitrarily close to the unique perturbation-free solution for small enough noise ). 

\section*{Acknowledgments}   
It is gratefully acknowledged the helpful comments of Andrew R. Teel, Luca Zaccarian, and Christophe Prieur.  

\bibliographystyle{elsarticle-num}
%\bibliography{Bibliography/biblio}
\bibliography{biblio}

\begin{thebibliography}{10}
\expandafter\ifx\csname url\endcsname\relax
  \def\url#1{\texttt{#1}}\fi
\expandafter\ifx\csname urlprefix\endcsname\relax\def\urlprefix{URL }\fi
\expandafter\ifx\csname href\endcsname\relax
  \def\href#1#2{#2} \def\path#1{#1}\fi

\bibitem{GSTbook}
R.~Goebel, R.~G. Sanfelice, A.~R. Teel, Hybrid Dynamical Systems: Modeling,
  Stability, and Robustness, Princeton University Press, 2012.

\bibitem{Schutter09}
B.~de~Schutter, W.~P. M.~H. Heemels, J.~Lunze, C.~Prieur, Survey of modeling,
  analysis, and control of hybrid systems, in: J.~Lunze, F.~Lamnabhi-Lagarrigue
  (Eds.), Handbook of Hybrid Systems Control, Cambridge University Press,
  Cambridge, 2009, pp. 31--55.

\bibitem{Goebel09}
R.~Goebel, R.~G. Sanfelice, A.~R. Teel, Hybrid dynamical dystems, IEE Control
  Systems Magazine 29 (2009) 28--93.

\bibitem{goebel06}
R.~Goebel, A.~R. Teel, Solutions to hybrid inclusions via set and graphical
  convergence with stability theory applications, Automatica 42~(4) (2006)
  573--587.

\bibitem{Prieur01}
C.~Prieur, Uniting local and global controllers with robustness to vanishing
  noise, Math. Control Signal Systems 14 (2001) 143--172.

\bibitem{PGT07}
C.~Prieur, R.~Goebel, A.~R. Teel, Hybrid feedback control and robust
  stabilization of nonlinear systems, IEEE Transactions on Automatic Control
  52~(11) (2007) 2103--2117.

\bibitem{LS99}
Y.~S. Ledyaev, E.~D. Sontag, A lyapunov characterization of robust
  stabilization, Nonlinear Analysis 37 (1999) 813--840.

\bibitem{copp16}
D.~A. Copp, R.~G. Sanfelice, A zero-crossing detection algorithm for robust
  simulation of hybrid systems jumping on surfaces, Simulation Modelling
  Practice and Theory 68 (2016) 1--17.

\bibitem{Horowitz75}
I.~M. Horowitz, P.~Rosenbaum, Nonlinear design for cost of feedback reduction
  in systems with large parameter uncertainty, International Journal of Control
  24 (1975) 977--1001.

\bibitem{libroResetBB}
A.~Ba{\~n}os, A.~Barreiro, {Reset Control Systems}, AIC Series, Springer,
  London, 2012.

\bibitem{Nesic11}
D.~Nesic, A.~R. Teel, L.~Zaccarian, Stability and performance of siso control
  systems with first order reset elements, IEEE Transactions on Automatic
  Control 56 (2011) 2567--2582.

\bibitem{Nesic05}
D.~Nesic, L.~Zaccarian, A.~R. Teel, Stability properties of reset systems, in:
  IFAC World Congress, Prague, Czech Republic, 2005.

\bibitem{Zaccarian11}
L.~Zaccarian, D.~Nesic, A.~R. Teel, {Analytical and numerical Lyapunov
  functions for SISO linear control systems with first-order reset elements},
  International Journal of Robust and Nonlinear Control 21 (2011) 71--76.

\bibitem{toolbox}
R.~G. Sanfelice, D.~A. Copp, P.~Nanez, {A toolbox for simulation of hybrid
  systems in Matlab/Simulink: Hybrid Equations (HyEQ) Toolbox}, in: Proceedings
  of the 16$^{th}$ international conference on Hybrid systems: computation and
  control, 2013, pp. 101--106.

\bibitem{Cai08}
C.~Cai, R.~Geobel, A.~R. Teel, Relaxation results for hybrid inclusions,
  Set-valued Analysis 16 (2008) 733--757.

\bibitem{Hahn67}
W.~Hahn, Stability of motion, Springer-Verlag, 1967.

\end{thebibliography}

\appendix
\small

\section{Hybrid systems solutions and basic properties (\cite{Goebel09,GSTbook,goebel06}))}
A subset $E$ of $\mathds{R}^n \times \mathds{N}$ is a {\em hybrid time domain} if it is the union of infinitely many intervals $[t_j,t_{j+1}]\times j$, where $0 = t_0 \leq t_1 \leq t_2 \leq \cdots$, or of finitely many such intervals, with the last one possibly of the form $[t_j,t_{j+1}]\times j$, $[t_j,t_{j+1})\times j$, or $[t_j,\infty)\times j$. A {\em hybrid arc} $\phi$ is a function $\phi:\text{dom } \phi \rightarrow \mathds{R}^n$, where  $\text{dom } \phi$ is a hybrid time domain and, for each $j$, $t \rightarrow \phi(t,j)$ is a locally absolutely continuous function on the interval
\begin{equation}
I_j = \{t:(t,j) \in \text{dom } \phi\}.
\end{equation}

The hybrid arc $\phi$ is a {\em solution to the hybrid system} $\mathcal{H} = (\mathcal{C},f,\mathcal{D},g)$ given by (1) (see \cite{Goebel09}) if $\phi(0,0) \in \mathcal{C}\cup \mathcal{D}$, and 
\begin{itemize}
\item (Flow condition) For each $j \in \mathds{N}$ such that $\text{int } I_j \neq \varnothing$,
\begin{equation}
\begin{array}{l}
\dot{\phi}(t,j)= f(\phi(t,j)), \text{  for almost all } t \in I_j, \\
\phi(t,j) \in \mathcal{C}, \text{  for all } t \in [\min I_j, \sup I_j),
\end{array}
\end{equation}
\item (Jump condition) For each $(t,j) \in \text{dom } \phi$ such that $(t,j+1) \in \text{dom } \phi$,
\begin{equation}
\begin{array}{l}
{\phi}(t,j+1)= g(\phi(t,j)), \\
\phi(t,j) \in \mathcal{D}.
\end{array}
\end{equation}
\end{itemize}
A solution $\phi$ to a hybrid system is {\em nontrivial} if $\text{dom } \phi$ contains at least one point different to $(0,0)$; {\em maximal} if it cannot be extended, that is there is no solution $\phi'$ with $\text{dom } \phi'$ contains $\text{dom } \phi$ as a proper subset, and such that $\phi'(t,j) = \phi(t,j)$ for any $(t,j) \in \text{dom } \phi$; and {\em complete} if $\text{dom } \phi$ is unbounded. 

There {\em exists nontrivial solutions} from $\xi \in \mathcal{C} \cup \mathcal{D}$ if there exist a discrete-time nontrivial solution or a continuous-time nonftrivial solution, that if either $\xi \in \mathcal{D}$ or there exist a solution $\mathbf{z}$ to $\dot{\mathbf{z}} = f(\mathbf{z})$ in some interval $[0,\epsilon]$, for some $\epsilon >0$, and satisfying $\mathbf{z}(0) = \xi$ and $\mathbf{z}(t) \in \mathcal{C}$ for $t \in[0,\epsilon]$. In addition, solutions are {\em unique} if and only if the following {\em basic uniqueness conditions} (see \cite{Goebel09}) hold: 
\begin{itemize}
\item for every $\xi \in \mathcal{C}\setminus \mathcal{D}$ there exists $\epsilon>0$ and a unique maximal solution ${\bf z}: [0,\epsilon] \rightarrow \mathds{R}^n$ to $\dot {\bf  z}(t)=f({\bf z}(t))$  satisfying ${\bf z}(0)=\xi$ and ${\bf z}(t)\in \mathcal{C}$ for all $t\in [0,\epsilon]$;
%\item For every $\xi \in \mathcal{D}$, $g(\xi)$ is a singleton
\item for every $\xi \in \mathcal{C} \cap \mathcal{D}$, there does not exist $\epsilon>0$ and an absolutely continuous ${\bf z} : [0,\epsilon] \rightarrow \mathds{R}^n$ such that $\mathbf{z}(0) = \xi$, $\dot{\mathbf{z}(t)} = f(\mathbf{z}(t))$ for almost all $t\in[0,\epsilon]$, and $\mathbf{z}(t) \in \mathcal{C}$ for all $t\in[0,\epsilon]$.
\end{itemize}

Given $T \geq 0$, $J \geq 0$, and $\varepsilon > 0$, {\em two hybrid arcs $\phi_1$ and $\phi_2$ are  $(T,J,\varepsilon)$-close} if: 
(a) for all $(t,j) \in \text{dom } \phi_1$ with $t \leq T$, $j \leq J$, there exists $s$ such that $(s,j) \in \text{dom }\phi_2$, $|t-s| < \varepsilon$, and $|\phi_1(t, j)-\phi_2(s, j)| < \varepsilon$; (b) for all $(t,j) \in \text{dom }\phi_2$ with $t \leq T$, $j \leq J$, there exists $s$ such that $(s,j) \in \text{dom }\phi_1$, $|t-s| < \varepsilon$, and $|\phi_2(t, j)-\phi_1(s, j)| < \varepsilon$.

%({\em hybrid system solution}, {\em forward completeness}, $(T,J,\epsilon)$-{\em closeness}, {\em outer perturbation}, etc) to be used in the following (see also Appendix).
%; and $\mathcal{S}_{\mathcal H}(K)$ is the set of maximal solutions $\phi$ to ${\mathcal H}$, with $\phi(0,0) \in K$

\section{Hybrid basic conditions (\cite{GSTbook,goebel06})}
A hybrid system with the data $(\mathcal{C},F,\mathcal{D},G)$ in $\mathds{R}^n$, satisfies the {\em hybrid basic conditions} if
\begin{enumerate}
\item $\mathcal{C}$ and $\mathcal{D}$ are closed sets.
\item $F:\mathds{R}^n \rightrightarrows \mathds{R}^n$ is outer semicontinuous and locally bounded, and $F(x)$ is nonempty and convex for all $x \in \mathcal{C}$.
\item $G:\mathds{R}^n \rightrightarrows \mathds{R}^n$ is outer semicontinuous and locally bounded, and $G(x)$ is nonempty for all $x \in \mathcal{D}$.
\end{enumerate}

For the hybrid system $\mathcal{H} = (\mathcal{C},f,\mathcal{D},g)$ given by \eqref{eq-H}, hybrid basic conditions are satisfied if $\mathcal{C}$ and $\mathcal{D}$ are closed sets, and $f$ and $g$ are continuous functions.% (\cite{GSTbook}).  

\textcolor{black}{
\section{Relaxation properties for hybrid inclusions}
For a hybrid system $\mathcal{H}$ with the data $(\mathcal{C},F,\mathcal{D},G)$ in $\mathds{R}^n$, 
$\mathcal{H}^\text{con}$ is defined by the relaxed hybrid inclusion
\begin{equation}
\mathcal{H}^\text{con}: \left\{  \begin{array}{llll}
\mathbf{\dot{x}} \in \text{con } F(\mathbf{x}), \quad & \mathbf{x}  \in \mathcal{C}, \\
\mathbf{x}^+ \in G(\mathbf{x}), \quad & \mathbf{x} \in \mathcal{D}.  
\end{array}
\right.
\label{eq-Hcon}
\end{equation}
Given $\mathbf{x}_0 \in \mathcal{C} \cup \mathcal{D}$, {\em strong relaxation} for all solutions from $\mathbf{x}_0$ is possible (\cite{Cai08}) if for any compact $\mathbf{x}: \text{dom }\mathbf{x}\rightarrow \mathds{R}^n$ with $\mathbf{x}(0, 0) = \mathbf{x}_0$ that is a solution to $\mathcal{H}^\text{con}$ and for any $\varepsilon >0$, there exists $\delta>0$ such that for any $\mathbf{y}_0 	\in (\mathbf{x}_0 + \delta \mathds{B}) \cap (\mathcal{C} \cup \mathcal{D})$ there exist a hybrid arc $\mathbf{y}:\text{dom }\mathbf{y} \rightarrow \mathds{R}^n$ with compact $\text{dom }\mathbf{y}$ and $\mathbf{y}(0, 0) = \mathbf{y}_0$ that is a solution to $\mathcal{H}$ and $d_\text{gph}(\mathbf{x},\mathbf{y}) \leq \varepsilon$, and moreover, if $\mathbf{x}(T, J ) \in \mathcal{D}$, where $(T, J ) = \max \text{ dom }\mathbf{x}$, then $\mathbf{y}(\tau, J ) \in \mathcal{D}$, where $(\tau, J ) = \max \text{ dom }\mathbf{y}$.
 }

 \textcolor{black}{
 In this work, it is used an extension of the strong relaxation property to cope with the problem of measurement noise and external disturbances. {\em Total strong relaxation} for all solutions from $\mathbf{x}_0$ is possible (\cite{Cai08}) if for any compact $\mathbf{x}: \text{dom }\mathbf{x}\rightarrow \mathds{R}^n$ with $\mathbf{x}(0, 0) = \mathbf{x}_0$ that is a solution to $\mathcal{H}^\text{con}$ and for any $\varepsilon >0$, there exists $\delta>0$ such that for any $\mathbf{y}_0 \in (\mathbf{x}_0 + \delta \mathds{B}) \cap (\mathcal{C} \cup \mathcal{D})$ and any admissible perturbations signals $\mathbf{n}_1$, $\mathbf{n_2}$, and $\mathbf{n_3}$ there exist a hybrid arc $\mathbf{y}:\text{dom }\mathbf{y} \rightarrow \mathds{R}^n$ with compact $\text{dom }\mathbf{y}$ and $\mathbf{y}(0, 0) = \mathbf{y}_0$ that is a solution to $\mathcal{H}^\text{con}_{\delta(\mathbf{n}_1,\mathbf{n_2},\mathbf{n_3})}$ and $d_\text{gph}(\mathbf{x},\mathbf{y}) \leq \varepsilon$, and moreover, if $\mathbf{x}(T, J ) \in \mathcal{D}$, where $(T, J ) = \max \text{ dom }\mathbf{x}$, then $\mathbf{y}(\tau, J ) \in \mathcal{D}$, where $(\tau, J ) = \max \text{ dom }\mathbf{y}$.
 }

\textcolor{black}{
Note that for $\mathcal{H} = (\mathcal{C},f,\mathcal{D},g)$, in which $f:\mathds{R}^n \rightarrow \mathds{R}^n$, simply  $\mathcal{H}^\text{con} =  \mathcal{H}$.
}

%For the hybrid system $\mathcal{H} = (\mathcal{C},f,\mathcal{D},g)$ given by \eqref{eq-H}, in the case in which $g$ is a continuous function and $\mathcal{C}$ and $\mathcal{D}$ are closed sets satisfying $\mathcal{C}\cup \mathcal{D}= \mathds{R}^n$ and $\text{int }\mathcal{C} \cap \text{int }\mathcal{D} = \varnothing$, the first hybrid relaxation condition is  directly satisfied. For the second relaxation condition, since $F = \text{con }F=f$ then \eqref{eqTHR-3a} and \eqref{eqTHR-3b} are given by
%\begin{equation}
%\dot{\mathbf{x}}(t) = f(\mathbf{x}(t)), \mathbf{x}(t) \in \mathcal{C} \text{ for almost all } t \in [0,T]
%\end{equation}
%and 
%\begin{equation}
%\dot{\mathbf{z}}(t) =  f(\mathbf{z}(t)+ \delta \mathbf{n}(t)), \mathbf{z}(t) \in \mathcal{C} \text{ for almost all } t \in [0,T]
%\end{equation}
%respectively.

\end{document}